\documentclass[11pt,a4paper]{article}

\pdfoutput=1
\usepackage[usenames,dvipsnames]{xcolor}
\usepackage{jheppubx}
\usepackage{enumitem}
\usepackage{graphicx}
\usepackage{dcolumn}
\usepackage{bm,psfrag}
\usepackage{overpic}

\usepackage[utf8]{inputenc}

\usepackage{amsmath}
\usepackage{cleveref}
\usepackage{subfigure}
\usepackage{slashed}
\usepackage{placeins}
\usepackage{float}
\usepackage[sf,isu]{caption}
\usepackage[mathscr]{euscript}

\usepackage{setspace}

\def\ket#1{|#1\rangle}
\def\Tr{{{\rm Tr}}}
\def\CO{{\cal O}}

\newcommand{\be}{\begin{equation}}
\newcommand{\ee}{\end{equation}}

\usepackage{chngcntr}

\makeatletter
\g@addto@macro\bfseries{\boldmath}
\makeatother

\newcommand{\bea}{\begin{eqnarray}}
\newcommand{\eea}{\end{eqnarray}}
\newcommand{\ba}{\begin{eqnarray}}
\newcommand{\ea}{\end{eqnarray}}
\newcommand{\nn}{\nonumber \\}

\newcommand{\beq}{\begin{equation}}
\newcommand{\eeq}{\end{equation}}
\newcommand{\beqa}{\begin{eqnarray}}
\newcommand{\eeqa}{\end{eqnarray}}
\newcommand{\beqar}{\begin{eqnarray*}}
	\newcommand{\eeqar}{\end{eqnarray*}}

\newcommand{\ie}{{\it i.e.,}\ }

\def\ausricht{\begin{aligned}}
	\def\endeausricht{\end{aligned}}

\def\t6 {T_\mt{D6}}

\newcommand{\mt}[1]{\textrm{\tiny #1}}

\def\sqr#1#2{{\vcenter{\vbox{\hrule height.#2pt
				\hbox{\vrule width.#2pt height#1pt \kern#1pt
					\vrule width.#2pt}\hrule height.#2pt}}}}

\usepackage{lipsum} 

\def\aa1{\phi}
\def\cc1{\psi}

\def\vev#1{\langle #1 \rangle}

\def\vev#1{\langle{#1}\rangle}

\def\makeitsmall{\begin{footnotesize}}
	\def\endmakeitsmall{\end{footnotesize}}

\def\nd{{ \vphantom{\dagger}}}

\def\nn{\nonumber}

\begin{document}

\title{Quantum quench, large $N$, and symmetry restoration }

\author{Diptarka Das and Bidyut Dey}
 \vspace{.1cm}

\affiliation{\vspace{0.2cm}
Department of Physics, 
Indian Institute of Technology - Kanpur, \\
Kanpur 208016, India.
\vspace{0.2cm}}

\date{\today}			

\emailAdd{$\lbrace$didas, bidyutd$\rbrace$@iitk.ac.in}

\abstract{We globally quench the theory of two dimensional massless fermions (\textit{many} flavours) with quartic interactions by making the quartic coupling a smooth function of time. Working in a derivative expansion we show that the discrete $\mathbb{Z}_2$ symmetry in case of the Gross-Neveu model, and the $U(1)$ symmetry in case of the Nambu-
Jona-Lasinio${}_{2}$ model, are restored during the zero-temperature quench. For the Gross-Neveu model we show that this can be understood as an effective thermalization. The time of symmetry restoration shows scaling with the quench rate. We identify this with the Kibble-Zurek scaling in the problem. In a suitable double scaling limit, the symmetry restoration may be understood in terms of Liouville quantum mechanics.  }

{
\hypersetup{urlcolor= blue!80!black}
\setlength{\parskip}{2pt}
\maketitle
\hypersetup{urlcolor=RoyalBlue!60!black}
}

\section{Introduction : Quench physics}

As most physics is out-of-equilibrium and tools are few, systems which allow analytical control are very valuable in understanding dynamics. It is thus not surprizing that a lot of effort has gone into studying quantum quenches in exactly and partially solvable systems. These include free quantum field theories \cite{Das:2016lla} where time-dependent solutions are exactly known. Next, there are quenches across quantum critical points where one can use conformal field theoretic technologies, which are especially powerful in 1+1 dimensions \cite{Calabrese:2005in}. There has also been a certain degree of progress in bosonic models with a large number of degree of freedom, inverse of which acts like an expansion parameter \cite{Sotiriadis:2010si}. Recently there has been progress in the field of sudden quenches for integrable theories, wherein the initial state ansatz can be bootstrapped using integrability techniques\cite{Bertini_2014}. The cited references serve as sample examples for each of the cases. One motivation of this work is to contribute to the above list and subsequently explore universalities that emerge in non-equilibrium dynamics.\\
An experimentally realizable out of equilibrium set-up is to drive the system by tuning the coupling ($g$) smoothly as a function of time. Even if we start from the ground state of the theory, the properties of the driven system now get contributions from \textit{excited states}. This is what makes the dynamics both difficult as well as interesting. Different universalities emerge depending upon the  details of the \textit{quench} protocol. Recent studies have shown that generically the  steady state turns out to have characteristics well described by a Gibbs or a Generalized Gibbs ensemble. When the quench is sudden then the analysis simplifies, since the problem now reduces to evolving an initially prepared state, $\ket{\psi_0}$ using a new post-quench Hamiltonian. When the quench is to a critical point, \cite{Calabrese:2005in} used techniques from boundary conformal field theory to investigate post-quench correlators and obtained universal features. The state $\ket{\psi_0}$ behaves like a thermal state with the temperature being related to the initial mass gap. Correlators calculated in this state  depend on the operator conformal dimensions. However, as all theories have a UV cut-off, smooth quenches are more realistic than sudden ones. The smoothness is characterized by a time scale $\delta t$ over which the change in coupling takes place, \ie, $g =g(t/\delta t )$. Various quench protocols are thus distinguished via various smooth functions. Such scenarios occur naturally, in the context of our expanding universe \cite{Boyanovsky:1996rw, Boyanovsky:1997cr}, as well as in heavy-ion-collisions at the RHIC experiments in LHC, \cite{Shuryak:2008eq, CasalderreySolana:2011us}. 
The latter phenomena is directly relevant towards understanding the steady state of the quark-gluon plasma. An important feature of the quark-gluon plasma is that it is strongly coupled. In this work we shall study smooth quenches in a 1+1 dimensional strongly interacting fermionic model that bears a lot of similarities with quantum chromodynamics, like asymptotic freedom, dimensional transmutation, and dynamical symmetry breaking.\\
Post-smooth-quenches universal features emerge depending on the rate of the quench. There are two main  regimes, the \textit{fast},  and the \textit{slow}. The fast is when $ \Lambda > \delta t^{-1} > E_g^{(0) }$, where $\Lambda$ is the ultra-violet (UV) cut-off and $E_g^{(0)}$ is the mass gap of the initial pre-quench state. In this case universal scaling emerges from the fixed point at the UV. This scaling has been observed in free theories, as well as using holography. See the dissertation  \cite{damianthesis} and references therein. The slow regime is when $\delta t^{-1} < E_g^{(0)} < \Lambda $. In this case too one may break adiabaticty. Obviously, there is no escape if one crosses a critical point during the quench, but adiabaticity also breaks if one comes \textit{close enough} to one. The criteria for adiabatic breakdown can be obtained by calculating the time dependent gap using \textit{adiabatic perturbation theory}. This is a calculation in a derivative expansion, where one assumes, $g > \dot{g} > \ddot {g} \dots$. It results in $E_g(t) = E_g^{(0)}(t) + E_g^{(1)}(t) + \dots$, where $E_g^{(1)}$ is the leading correction to the gap and further corrections are denoted by the $\dots$. This expansion breaks down when $E_g^{(1)} \sim E_g^{(0)}$, which happens at a particular time, dubbed as the Kibble-Zurek (KZ) time, $t_{KZ}$. $t_{KZ}$ is the time when the system fails to respond to the quench, and all correlators freeze at this scale. The KZ time is generically a function of the rate, $t_{KZ} = t_{KZ}(\delta t)$. Operator expectations are dictated by this scale : $\vev{\CO_\Delta}\sim t_{KZ}^{-\Delta}$ \cite{Kibble:1976sj, Zurek:1985qw}. For a nice review including experimental results, see \cite{Polkovnikov:2010yn}. Both the \textit{fast} as well as the KZ scalings have extensively been investigated in explicitly solvable models either in the continuum or on the lattice. See \cite{Das:2016eao, Das:2014hqa, Das:2014jna, Das:2015jka, Das:2016lla} for studies in the continuum and \cite{Mondal2010, Divakaran2010, Das:2017sgp} for some lattice examples. In all of the field theoretic examples exhibiting the KZ scalings, investigations have been limited to Hamiltonia which can be quadraticized, else in scenarios involving one ( or more ) critical point(s), and thereby in certain cases using the CFT technology. It is a much wider and unknown area to explore quenches in interacting quantum field theories away from criticality. Recently, in \cite{Goykhman:2018iaz} the authors considered a $\phi^4$ theory in $4-\epsilon$ dimensions and addressed the quench of the quartic coupling constant. Perturbatively (in the quartic coupling) they were able to show how renormalization group effects via time-dependent counterterms,  induce a quench in the running mass. In the quartic fermionic model we also find a induced time dependent gap resulting from the quartic coupling quench. Our calculation however is non-perturbative in the coupling, as we rely on the largeness of the number ($N$) of fermionic flavours, which allows us to focus on a special class of diagrams that can be summed. \\
Large $N$ methods have previously been used to study quenches in interacting theories \cite{Boyanovsky:1997cr, Boyanovsky:1996rw, Sotiriadis:2010si, Das2012, Gemsheim:2019qed}. As a result of the quench the system often exhibits an effective thermal behaviour, however there has not been any explicit KZ scaling analysis in this setting. In our work we present such a scenario where a KZ scale emerges. We find that the scaling is \textit{tied} to restoration of a dynamically broken symmetry of the system. It is also known that this broken symmetry gets restored at finite temperatures \cite{Jacobs:1974ys, Harrington:1974tf}. We show that the restoration during the quench can be understood as an \textit{effective thermalization} at a temperature which takes the system to the symmetric, disordered phase. While one is guaranteed to break adiabaticity while crossing a critical point, the adiabatic expansion can still breakdown \textit{close} to the critical point. In our set-up too we shall stay close to the $g=0$, UV fixed point, close enough in order to get the emergent KZ scaling.\\
\textbf{Organization :} In \S\ref{sec:GN} we describe the model and briefly state its equilibrium properties. Then we set up the problem in the Schwinger-Keldysh contour which allows us to conveniently study the quench in large $N$. In \S\ref{sec:dynsad} we focus on the dynamical gap equation using derivative expansion. In \S\ref{sec:num} we solve it numerically and state the results which show that quenches lead towards restoration of a discrete symmetry. In \S\ref{efftherm} we argue how the broken-restoration transition during quench can be interpreted as an effective thermalization. Next we investigate KZ in \S\ref{sec:KZ} and analytically argue for the scaling of the symmetry restoration time. In the subsection \S\ref{sec:liou} we show that the order parameter dynamics has a natural double scaling limit wherein the symmetry restored configuration can be understood via Liouville quantum mechanics. We outline some future directions and end with some discussions in \S\ref{sec:conc}. Appendix \S\ref{app:eps} contains detail of the derivative expansion. Numerics relevant for \S\ref{sec:KZ} is relegated to appendix \S\ref{app:tanhnum}. A full-fledged analysis of a similar model is carried out in \S\ref{app:njl} which also exhibits restoration of symmetry, albeit a continuous one.

\section{Setting-up the quench in the Gross-Neveu model}\label{sec:GN}
The focus of our study is the theory of two dimensional massless fermions with quartic interactions, introduced by Gross and Neveu \cite{PhysRevD.10.3235}. The Gross-Neveu model is a renormalizable field theory admitting a $1/N$ expansion (where $N$ is the number of fermion flavours) and displays asymptotic freedom and dynamical symmetry breakdown. The discrete chiral symmetry $\mathbb{Z}_2$ of the model is dynamically broken via the generation of a $\langle\bar{\psi} \psi \rangle$ mass term, which is non-perturbative in the quartic coupling. It is also known that the symmetry gets restored at finite temperature \cite{Jacobs:1974ys, Harrington:1974tf} and also in presence of finite curvature \cite{buchkiri89}. We implement the smooth quench by promoting the constant quartic coupling $g$ to $g(t/\delta t)$. A global quench can be thought of as turning on a time-dependent metric. Furthermore, it is also expected that a strongly interacting system will eventually thermalize. Therefore it may be expected that under a quench the symmetry may get restored eventually.   
At equilibrium, the Gross-Neveu (GN) action is given by,
\begin{equation}
S =\int d^2 x \{ \bar{\psi}_{i} i \slashed{\partial} \psi_{i} + \frac{1}{2N}g^2 (\bar{\psi}_{i} \psi_{i})^2\} \label{eq:a}
\end{equation}
where coupling $ g^2 $ is kept fixed as $ N \rightarrow \infty $. The large number $N$ here, is the number of fermionic species, $i \in \{1, N\}$. In two dimensions the bare fermion mass dimension is $\tfrac{1}{2}$, hence the coupling is dimensionless and consequently gets only logarithmic corrections $\implies$ the theory is renormalizable. $\gamma^0$ and $\gamma^1$ are $2\times 2$ gamma matrices in 2 dimensions, which we take to be in the Weyl basis (\S \ref{free}). A mass term is a priori excluded by the discrete chiral symmetry,
$$
\psi_i \rightarrow \gamma^5 \psi_i, \,\, \bar{\psi}_i \rightarrow - \bar{\psi}_i \gamma^5. $$ It is this symmetry that suffers from a dynamic breakdown. To solve the theory in large $N$ one introduces an auxiliary scalar field $\sigma$, 
\begin{equation}
S = \int d^2 x \{ \bar{\psi} i \slashed{\partial} \psi -\frac{N}{2 g^2} \sigma^2 + \sigma
\bar{\psi} \psi   \} .\label{eq:b}
\end{equation}
Integrating over $\sigma$ gives back eq\eqref{eq:a}. The discrete chiral symmetry acts as a simple $\mathbb{Z}_2$,  $\sigma \rightarrow - \sigma$. In equilibrium one can regard $\sigma(x,t) = \sigma$ to be a spacetime constant. Integrating out the fermions and then evaluating the saddle in the $\sigma$ functional integral yields, \begin{equation}\label{equilibrium} \sigma  = 2\Lambda \frac{e^{\pi/g^2}}{e^{2\pi/g^2}-1} \sim 2\Lambda e^{- \pi/g^2 }, \end{equation} where $\Lambda$ is the cut-off scale. Note that the last $\sim$ approximation holds for $g\rightarrow 0$ only. In our analysis we do not make any small $g$ approximation. This non-zero value of $\sigma = \tfrac{g^2}{N} \bar{\psi}\psi$ signals the breakdown of the $\mathbb{Z}_2$ symmetry, and results in a mass for the fermions. For the time-dependent $g$, all amplitudes now need to be calculated using the Schwinger-Keldysh contour. Also as we shall see, that we can no longer choose $\sigma(x,t)$ as a spacetime constant. The partition function evaluated using the Schwinger-Keldysh contour is given by, 
\begin{eqnarray}
\mathcal{Z} = \int \mathcal{D} \bar{\psi}_{\pm} \mathcal{D}\psi_{\pm} \mathcal{D}\sigma_{\pm} \exp \left[ i \{S(\bar{\psi}_{+},\psi_{+}, \sigma_{+}) - S(\bar{\psi}_{-},\psi_{-}, \sigma_{-}) \} \right] \label{eq:e}
\end{eqnarray}
where,
\begin{align}
S(\bar{\psi}_{+},\psi_{+}, \sigma_{+}) - S(\bar{\psi}_{-},\psi_{-}, \sigma_{-}) &=  \int d^2 x [\bar{\psi}_{+} ( i  \slashed{\partial}  + \sigma_{+} ) \psi_{+} - \bar{\psi}_{-} ( i \slashed{\partial} + \sigma_{-} ) \psi_{-} \nn \\&- \frac{N}{2 g^2(t)} (\sigma^2_{+} - \sigma^2_{-})  ] \label{eq:f}.
\end{align}
Next, we integrate out $ \bar{\psi}_{\pm}, \psi_{\pm}$ to obtain the effective action,
\begin{align}
&S_{\text{eff}}(\sigma) = -N \, \Tr \log D - \frac{N}{2g^2(t)} \int d^2 x (\sigma_{+}^2 - \sigma_{-}^2) \label{eq:g}, \nn \\&\text{ with}
\,\,\, D= \left( \begin{matrix}
i\slashed{\partial} + \sigma_{+} && 0 \\
0 && - i\slashed{\partial} - \sigma_{-}
\end{matrix} \right).
\end{align}
At large-$N$, $S_{\text{eff}}(\sigma)$ is dominated by the saddle point configuration which is given by, 
 \begin{equation}\label{saddle}
\frac{\sigma_{\pm}}{g^2(t)} = - \Tr\left[\frac{1}{ i \slashed{\partial} + \sigma_{\pm}} \right] = - \int_{-\infty}^\infty dk\, \langle \bar{\chi}_k(t) \chi
_k(t) \rangle. 
\end{equation}
Note, that in general due to the boundary condition at the turning point of the Schwinger-Keldysh contour, there are non-trivial correlations between the $+$ and the $-$ fields, however the classical saddle remains unaffected \cite{Kamenev_2009}. The first equality in eq\eqref{saddle} shows that $\sigma$ can no longer be constant in time. Therefore in the second equality, $ \chi$ is a free massive fermion in two spacetime dimensions with a time dependent mass $m(t)= -\sigma(t)$, whose correlator we seek. The fermionic field $\chi$ is used as an auxiliary field employed to calculate the trace. This is the problem we turn to next.
\subsection{Free fermion with time-dependent mass}\label{free}
The relevant Dirac equation that we solve for is,
\begin{equation}
\left( i \gamma^0 \partial_0 + i \gamma^1 \partial_1 - m(t) \right) \chi(x,t) = 0. 
\end{equation}
We follow the conventions used in \cite{PhysRevD.97.125012}, and work in the Weyl-basis where the $\gamma$ matrices take the form,
\begin{align} \label{weylbasis}
\gamma^0 &= \begin{pmatrix} 0 & 1 \\ 1 & 0 \end{pmatrix}, \,\,\,\,\,\gamma^1 =  \begin{pmatrix} 0 & 1 \\ -1 & 0 \end{pmatrix}. 
\end{align}
Next, we decompose the Dirac field into momenta modes,
$\chi(x,t) = \sum_k e^{i k x} \chi_k (t) . $
The equation satisfied by $\chi_k(t)$ is, 
$
\left( i \gamma^0 \partial_0 -k  \gamma^1  - m(t) \right) \chi_k(t) = 0. 
$
We choose the spinor basis of solutions as,
\begin{align}
u_k(t) &= \frac{ e^{ikx} }{\sqrt{2\pi }  } \begin{pmatrix} h_k^I(t) \\ -h_k^{II}(t) \end{pmatrix} , \,\,\,
v_k(t) = \frac{ e^{-ikx} }{\sqrt{2\pi }  } \begin{pmatrix} h_{-k}^{II*}(t) \\ h_{-k}^{I*}(t) \end{pmatrix}.
\end{align}
The equation of motion in momentum space translates into the following coupled equations of motion, 
\begin{align} \label{h1h2}
\dot{h}_k^I - i k h_k^I - i m h_k^{II} &= 0, \,\,\, 
\dot{h}_k^{II} + i k h_k^{II} - i m h_k^{I} = 0,
\end{align}
along with the standard normalization condition, 
$
| h_{\pm k}^I |^2 + | h_{\pm k}^{II} |^2 = 1. 
$
Now we can write down the second-quantized mode expansion for the Dirac field as,
$
\chi(x,t ) = \int\, dk \,\, [ B_k^\nd u_k(t,x) + D^\dagger_k v_k(t,x) ],
$
where, 
 $\{ B_k^\nd  , B_{k'}^\dagger\} = \delta(k-k')$ and   $\{ D_k^\nd , D_{k'}^\dagger\} = \delta(k-k')$. 
With these relations one can check that, 
$ \{ \chi^\nd(x,t) , \chi^\dagger(y,t) \}$ $= \delta(x-y) \mathbb{I}_{2\times 2}. 
$
A straightforward algebra gives, 
$$
\int dk\, \langle \bar{\chi}_k(t) \chi_k(t) \rangle = \int_{-\infty}^\infty \frac{dk}{2\pi} \left( h_{-k}^I h_{-k}^{II*} + h_{-k}^{II} h_{-k}^{I*} \right
).$$ To proceed we propose the following \textit{ansatz}, 
\bea
h_k^I &=& \sqrt{ \frac{\omega_k(t) - k }{2\omega_k(t) }} e^{-i \int^t \omega_k(t') dt' },\,\,\,
 h_k^{II} = -\sqrt{ \frac{\omega_k(t) + k }{2\omega_k(t) } }e^{-i \int^t \omega_k(t') dt' }. 
 \label{ansatz}
 \eea
This is inspired from equations \eqref{h1h2}\footnote{ In equilibrium, the ansatz automatically satisfies, $h_k^{II} = \tfrac{1}{im} \dot{h}_k^I - \tfrac{k}{m} h_k^I$, and maintains $
| h_{\pm k}^I |^2 + | h_{\pm k}^{II} |^2 = 1. 
$ for all times.}. Note, that this is quite different from the ansatz used in the bosonic cases \cite{Gemsheim:2019qed, Das2012}, wherein there is only a single mode to solve for, and the prefactor is also considerably simpler. 

\section{Dynamic saddle : symmetry restoration}\label{sec:dynsad}
With the ansatz \eqref{ansatz}, the gap equation \eqref{saddle} simplifies to, 
\begin{equation}
\frac{\sigma(t) }{g^2(t)} = \int_0^\infty \frac{dk}{\pi } \mathcal{A}(k,t),\,\,\text{where,}\, \mathcal{A}(k,t)= \frac{ \text{Re} \bigg[ \sqrt{\omega_k(t) - k } \bigg( \sqrt{ \omega_k(t) + k }\bigg)^* \bigg]}{ | \omega_k(t) | }. \label{saddle2}
\end{equation}
Note, that in the static equilibrium case, the above reproduces the correct gap equation, 
\begin{equation}
\frac{1}{g^2} =  \int_0^\infty \frac{dk}{\pi } \frac{1}{\sqrt{ k^2 + \sigma^2 } },
\end{equation}
which has the equilibrium solution, given by 
\begin{equation} \label{equi}
\sigma = m_0= 2 \Lambda \frac{ e^{\pi / g^2}}{ e^{2\pi / g^2 }  -1 }.
\end{equation}
 One can now use the  derivative expansion to solve the for $\omega_k(t)$. The differential equation satisfied by $\omega_k(t)$ can be obtained from eq\eqref{h1h2} by plugging in the ansatz eq\eqref{ansatz} (eq\eqref{ode1}). See \S Appendix \ref{app:eps} for further details. The integrand in eq\eqref{saddle2} takes the form given by equation \eqref{Akt}. Integrating this over the momentum (with cut-off $= \Lambda$) we  obtain a second order differential equation for the saddle $\sigma(t)$,
\begin{align}\label{ode2}
&\frac{24 \pi \sigma^4 }{g^2}= 6  ( 4\sigma^4 - \dot\sigma^2 +\sigma \ddot\sigma ) \log \bigg( \Lambda + \sqrt{\Lambda^2 + \sigma^2 } \bigg) + \frac{3 \sigma^2 ( \dot\sigma^2 + \sigma \ddot\sigma )}{\Lambda^2 + \sigma^2 } \nn \\ &+ \frac{2\Lambda \sigma^2 ( 2\dot\sigma^2 + \sigma \ddot\sigma )}{( \Lambda^2 + \sigma^2 )^{3/2} }+ \Lambda \frac{5\dot \sigma^2 - 8 \sigma \ddot \sigma }{\sqrt{\Lambda^2 + \sigma^2 } } - 3( \dot \sigma^2 - \sigma \ddot \sigma ) \log \bigg( \Lambda^2 + \sigma^2 \bigg)  - 3\sigma^4 \dot\sigma^2  \nn \\ &\bigg( \frac{\Lambda}{(\Lambda^2 +\sigma^2 )^{5/2} } + \frac{1}{(\Lambda^2 +\sigma^2 )^2 } \bigg)
- 3 \sigma \ddot\sigma (1 + 4\log \sigma ) + 12 \dot\sigma^2 \log \sigma - 24 \sigma^4 \log \sigma. 
 \end{align}
We solve this equation numerically (without any approximation) with initial equilibrium conditions that depend upon the quench protocol. Fig.\ref{b5der} justifies the derivative expansion, in that the solution $\sigma(t)$ is always larger than its time derivatives. 

\subsection{Numerical results}\label{sec:num}

We use $\texttt{Mathematica}^{\tiny{\textregistered}}$ to numerically integrate the differential equation \eqref{ode2} for  the smooth standard quench profile asymptoting between two constant values, the $\tanh t/\delta t$ function. Working within the regime of validity of the approximation, we find that the order parameter quickly settles to zero, signalling the approach towards \textit{restoration} of the dynamically broken symmetry. We choose equilibrium initial conditions at early times, \ie equation \eqref{equi} along with $\dot\sigma(-\infty) = 0$. 

\subsubsection*{Tanh quench}
\begin{figure}[h!]
        \centering
                \includegraphics[scale=0.52]{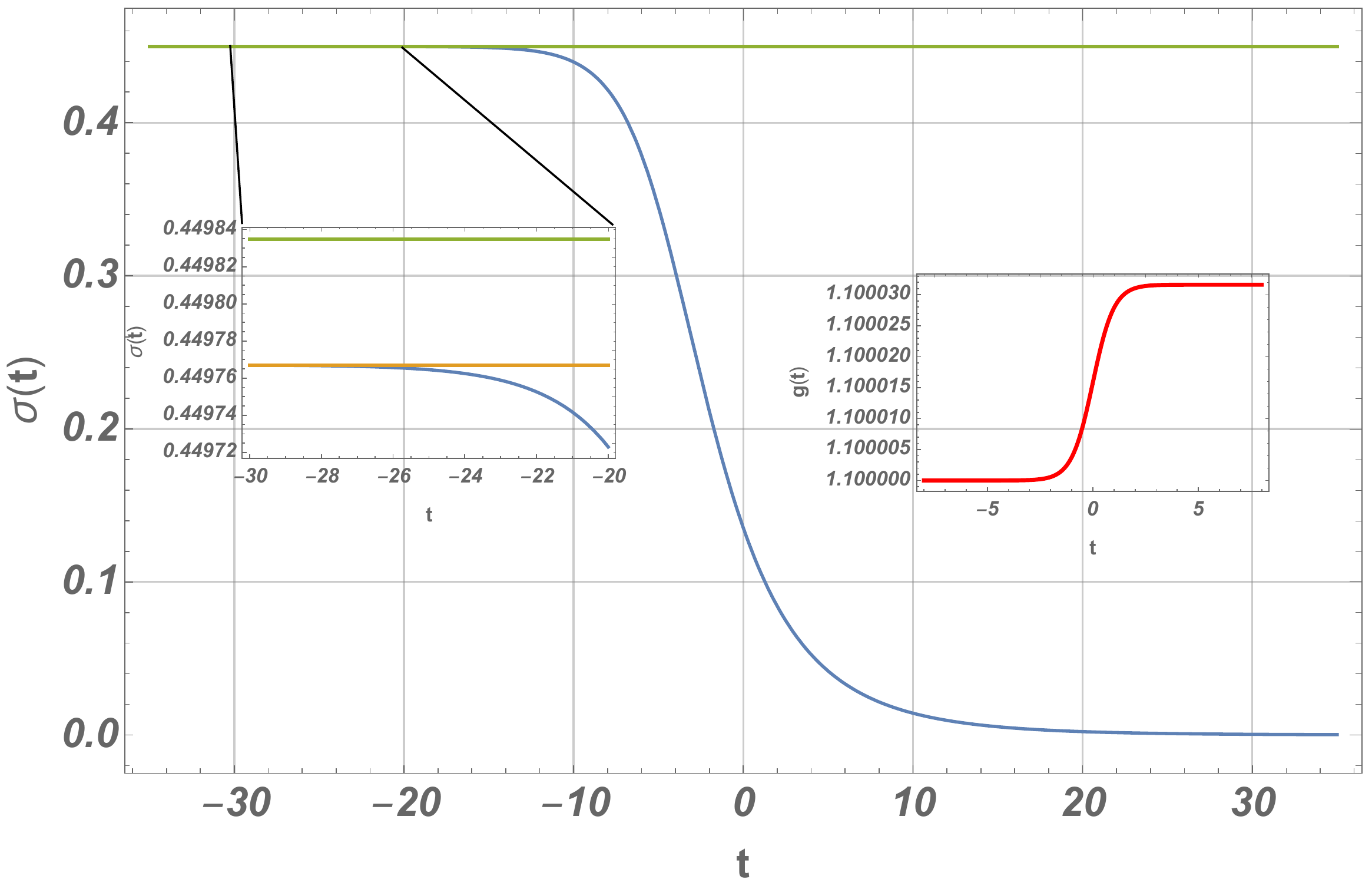} 
                 \caption{ Plot of $\sigma(t)$ as a function of time. We have chosen cut-off, $\Lambda = 3$ with $g_i = 1.10, g_f = 1.1 + 10^{-4.5}, \delta t = 10$, the red inset curve shows $g(t)$ vs. $t$. In the zoomed inset, the orange line corresponds to the equilibrium $\sigma$ for $g=g_i$ and the green corresponds to $\sigma$ for $g=g_f$. }
                 \label{tanh1}
\end{figure}

More explicitly we choose, 
\begin{equation}\label{tanhprof}g(t) = \frac{ g_i + g_f}{2} + \frac{g_f - g_i}{2} \tanh \frac{t}{\delta t}. \end{equation}
The plot in Fig. \ref{tanh1} shows how the order parameter quickly settles to vanishing value. 

\FloatBarrier

\subsubsection{Breaking-Restoring transition}\label{brt}
\begin{figure}[!htbp]
        \centering
        \subfigure[Small amplitude Tanh quench]{
                \includegraphics[scale=0.4]{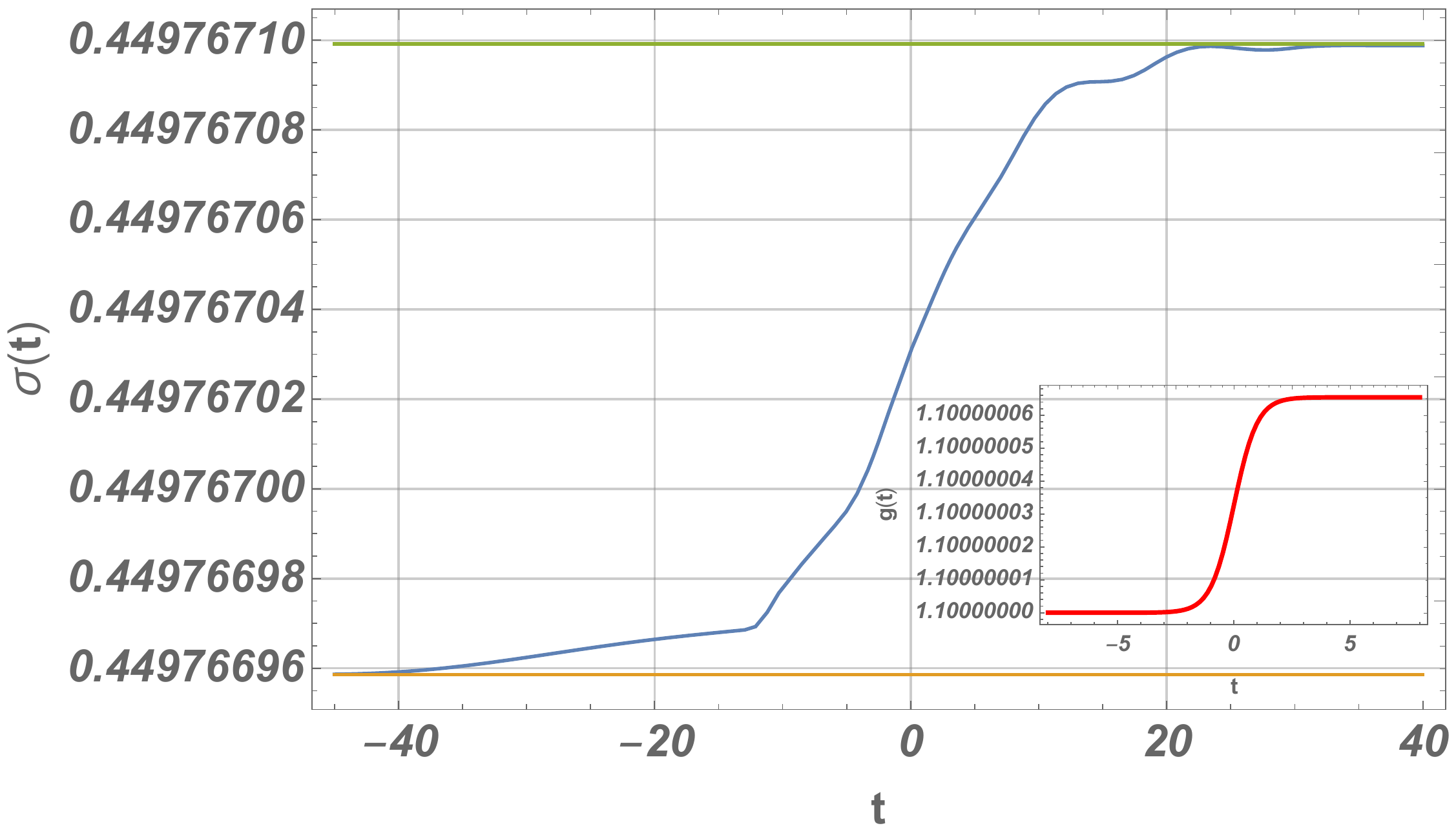} \label{tanhsmall}}
         \subfigure[Transition from broken to restoration]{
                \includegraphics[scale=0.47]{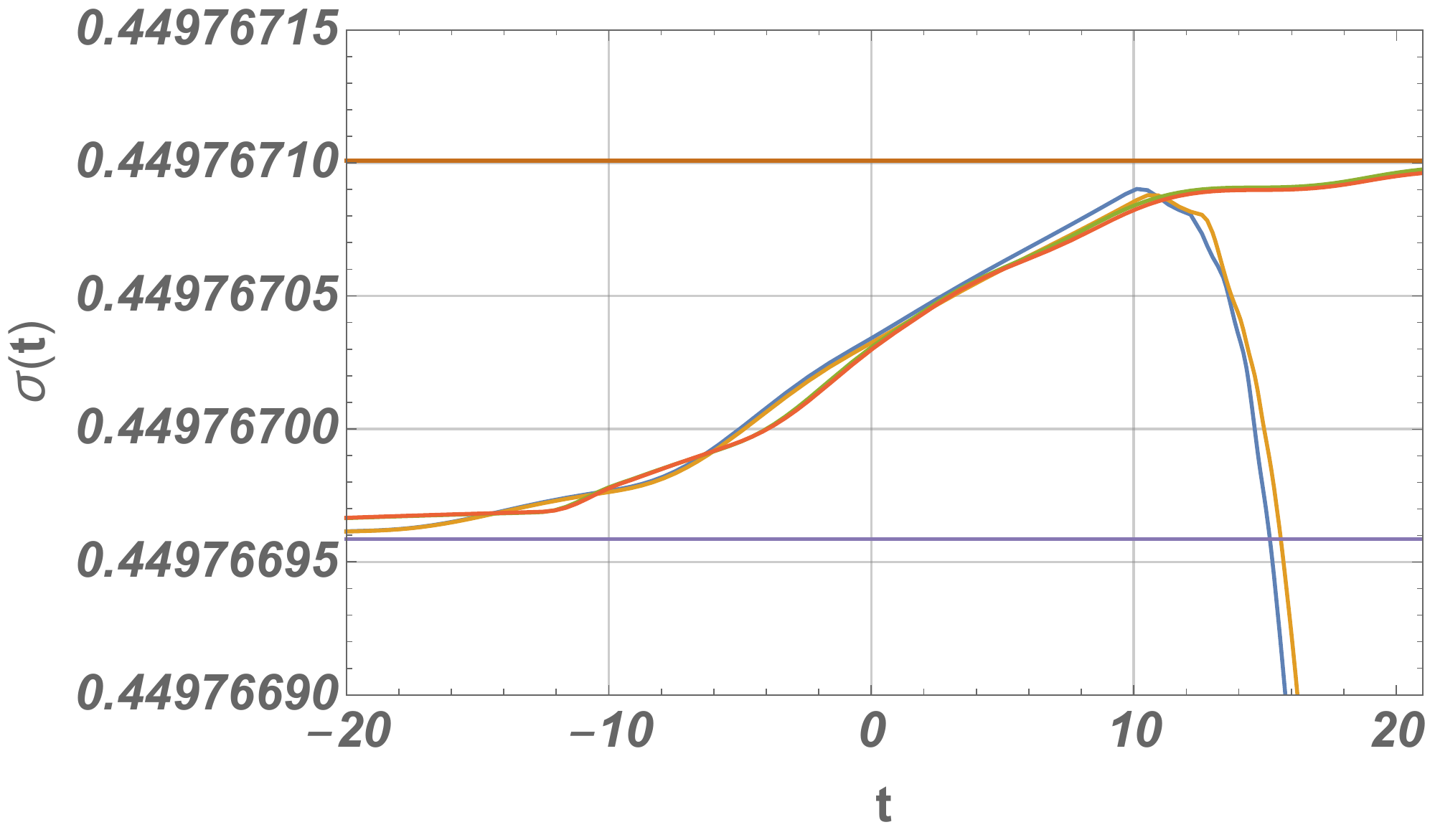} \label{tanhsmallbig}}
                 \caption{(a) Plot of $\sigma(t)$ as a function of time. We have chosen, $g_i = 1.1, g_f = 1.1+10^{-7.184}, \delta t = 10, \Lambda = 3$, the red inset curve shows $g(t)$ vs. $t$. The orange line corresponds to the equilibrium $\sigma$ for $g=g_i$ and the green corresponds to $\sigma$ for $g=g_f$. In (b) we plot the various $\sigma$ profiles for different quench amplitudes. }
\end{figure}

Here we numerically investigate for the \textit{tanh} quench protocol, the transition from the broken phase to the restored phase as we vary the quench amplitude for a fixed $\delta t$. We notice from Fig.(\ref{tanhsmall}) that if $g_i$ and $g_f$ are very close to each other, then the quench does not restore the broken symmetry. In fact, there is a transition at a particular quench amplitude. In Fig.(\ref{tanhsmallbig}) we show the evidence for the transition. In this figure, the green and the red quenches do not restore the broken symmetry, while the orange and blue quenches do\footnote{One can once again check the self-consistency of the numerical solutions by comparing with the time derivatives as carried out for Fig.(\ref{b5der}).}. The green and red  corresponds to the case, $g_f-g_i < 10^{-7.18}$ while the orange and the blue curves corresponds to $g_f-g_i > 10^{-7.18}$. The critical amplitude depends on the rate $\delta t$. A slower quench requires larger quench amplitude to restore the broken symmetry. It is the amplitude of the quench at a fixed rate which causes the transition. One could have equivalently chosen $g_i > g_f$ and also discussed this transition.\\
In the next section we have shown that  this dynamic transition for a fixed rate as a function of the quench amplitude can be phrased in terms of thermalization. Later we have also focussed on the dependence of the transition time on the rate of the quench at a fixed amplitude. It is this latter case that can be associated to a Kibble-Zurek scaling. 

\FloatBarrier

\subsubsection{Effective thermalization}\label{efftherm} 
\begin{figure}[!htbp]
        \centering
         \includegraphics[scale=0.45]{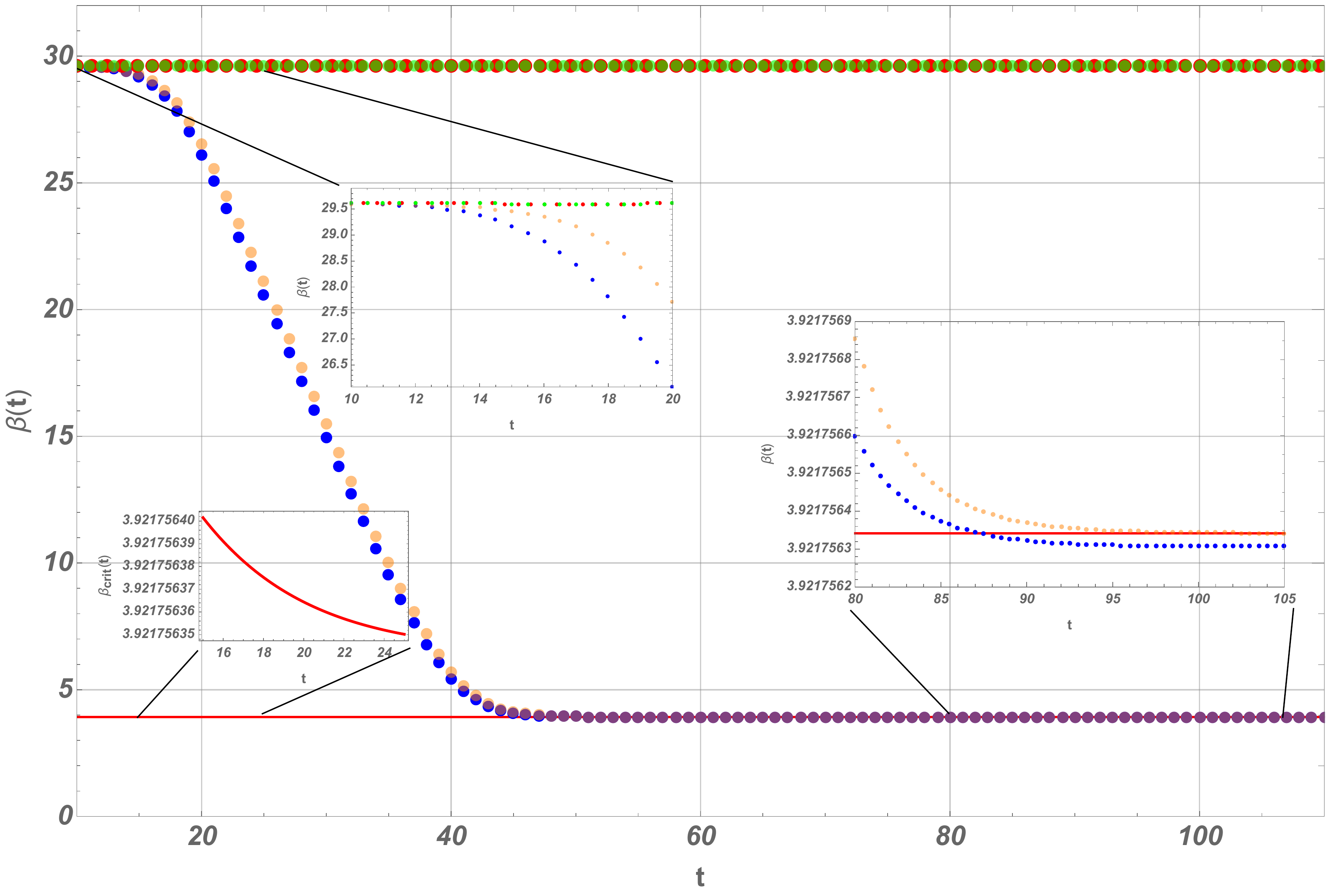} 
         \caption{ Plot of effective temperature as a function of time for various $tanh$ quench profiles. We have chosen, $\Lambda =3$, $g_i = 1.1$, $\delta t = 10$. The blue, orange, red and green corresponds to choosing $g_f = \{1.1+10^{-7.16},1.1+10^{-7.172},1.1+10^{-7.184},1.1+10^{-7.188} \}$, respectively. The insets are zoomed in views of different parts. }
         \label{tanh-beta}
\end{figure}

In this subsection we show how the symmetry broken-restoration transition during the quench can be understood as an effective thermalization. It is useful firstly to define an \textit{instantaneous gap}, $\text{m}(t) = 2\Lambda \tfrac{e^{\pi /g^2(t)}}{e^{2 \pi /g^2(t)} -1 }$ which becomes the gap in equilibrium, $m_0$, eq\eqref{equi}. Unlike the non-linear bosonic O(N)${}_3$ model where the thermal phase transition can be analytically investigated \cite{sachdev_2011}, the GN${}_2$ thermal transition can only be fully understood numerically \cite{Jacobs:1974ys, Harrington:1974tf}. In equilibrium as the temperature is increased the $\sigma =0$ point becomes the true minima instead of $m_0$ and the chiral symmetry gets restored. The critical inverse temperature is found to be, $\beta_{\text{crit}} = \frac{\pi e^{-\gamma} }{m_0}$. In the quench context we therefore naturally define a critical time-dependent $\beta_{\text{crit}}(t) = \frac{\pi e^{-\gamma} }{m(t)}$. In equilibrium, one can also define a temperature-dependent fermionic mass by extremizing the finite-temperature effective potential\cite{Jacobs:1974ys}. This is given by, 
\begin{equation}\label{temp}
    2 f\left( \beta^2 m_\beta^2 \right) = \gamma + \log \frac{m_0 \beta}{\pi}. 
\end{equation}
In the above equation $$f(a) = \displaystyle \sum_{n=0}^\infty \frac{1}{2n+1}\left\{ 1 - \left( 1 + \frac{a}{(2n+1)^2\pi^2}\right)^{-1/2} \right\},$$ and $\gamma= 0.5772156649$ is the Euler-Mascheroni constant. Unfortunately, eq\eqref{temp} cannot be solved to obtain, $m_\beta$ as a function of $\beta$ or vice-versa in a closed form. However from the time-dependent analog of eq\eqref{temp} we can numerically extract an \textit{effective} $\beta(t)$. The equation we solve is, 
\begin{equation}\label{temp2} 2 f\left(\beta^2(t) \sigma^2(t)\right) = \gamma + \log \tfrac{m(t) \beta(t) }{\pi}. \end{equation}
We plot the results in Fig.\ref{tanh-beta}. We choose parameters and colours as in \S\ref{brt}. We notice that both the symmetry restoring quenches (orange, blue points) cross  $\beta_{\text{crit}}(t)$ (red curve) while the symmetry broken quenches (red and the green points) do not!

\FloatBarrier

\section{Emergence of Kibble-Zurek scaling} \label{sec:KZ}
\begin{figure}[!htbp]
\centering
\includegraphics[scale=0.5]{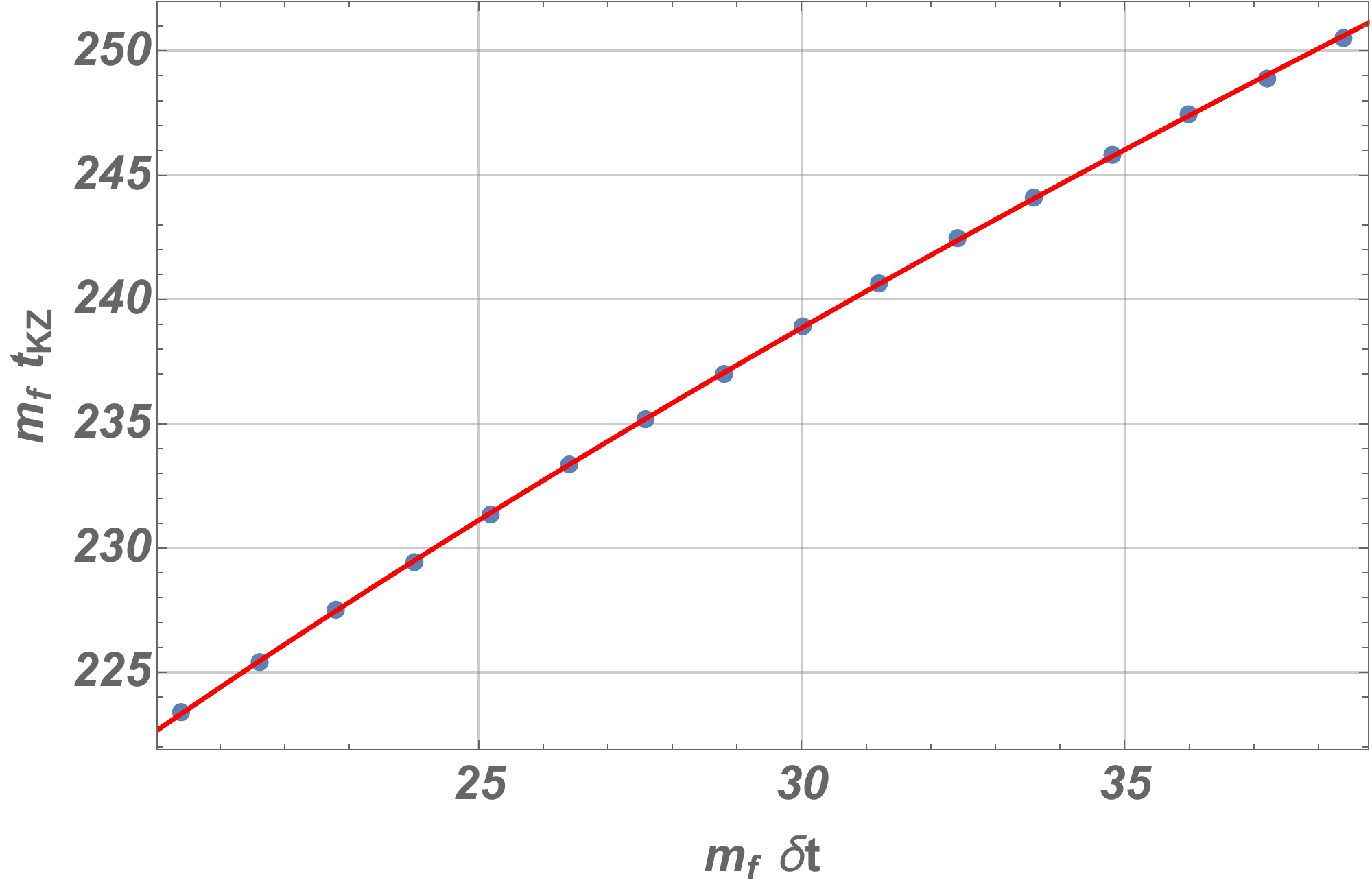} 
\caption{ A fit of the zero times extracted for various ranges as in Fig.\ref{tzero} with the scaling function given by eq\eqref{scaling}. We use parameters as listed in eq\eqref{params}.}\label{tkz}
 \end{figure}

In this section, we show that the time at which the order parameter goes to zero, can be identified with the Kibble-Zurek (KZ) time scale in the problem. The Kibble-Zurek time is the moment in time when the adiabatic corrections to the time-dependent gap\footnote{which depends non-trivially on the order parameter and its derivatives.}, become of the same order as the leading answer. This emergent time-scale dictates all correlations in the system, which freezes at the KZ scale. As in \S\ref{efftherm}, it is useful to think of the quench as a change in the dimensionful  parameter, $m(t)$, which is identified with the mass gap at equilibrium. We let $m(t) =  \frac{ m_i + m_f}{2} + \frac{m_f - m_i}{2} \tanh \frac{t}{\delta t}$. This is achieved practically by making $g(t)$ a function of time. Next, we plot the dynamic order parameter $\sigma(t)$ during the quench for different $\delta t$ values while keeping the amplitude of the quench fixed. The results are consistent with the expectation that for slower quenches the order parameter takes longer time to vanish. See appendix \S\ref{app:tanhnum} for further details. We have taken  zero to be some fixed small number ($\epsilon$) and find the set of times when $\sigma = \epsilon$ for the different $\delta t$'s. We make sure to be in the regime, $\Lambda > m_f, \, m_i > \delta t^{-1}$, where one expects KZ. However as emphasised in the introduction, we do not cross criticality (\ie $\,$vanishing $m(t)$). Now, we  find analytically the scaling of $t_{KZ}$ with $\delta t$ and show that the zero times extracted from the numerics exhibit the same KZ scaling. \\
The analytical analysis is simpler in the $\Lambda/\sigma \gg 1$ limit of \eqref{ode2}  which simplifies now to, 
\begin{equation}\label{limit}
\frac{\pi}{g^2} = \log \frac{\Lambda}{\sigma} - \frac{ \dot\sigma^2}{2\sigma^4} \log \frac{\Lambda}{\sigma} + \frac{\ddot \sigma}{2\sigma^3} \log \frac{\Lambda}{\sigma}.
\end{equation}
In the derivative expansion this equation can now be inverted to give (till quadratic order),
\begin{equation}
\sigma = m \left( 1 + \frac{\pi}{2 m^2 g^2 }\left( \frac{\ddot{m}}{m} - \frac{\dot{m}^2 }{ m^2 } \right) + \dots \right).
\end{equation}
The KZ scale in the problem, arises from the condition of adiabatic breakdown near criticality when the equilibrium mass $m(t) \rightarrow 0$. Mathematically, this occurs when the correction term is of the leading order. Near this region for slow rates,  assuming $m(t) \sim t/\delta t$, we find the adiabaticity breakdown condition to be, 
$ \log t_{KZ}/\delta t \simeq t_{KZ}^4/\delta t^2$, this can be solved for $t_{KZ}$ in terms of the Lambert function to give the scaling: 
\begin{equation}\label{scaling}
t_{KZ} \sim \sqrt{\delta t} \,\, W(4\delta t^2)^{1/4}. 
\end{equation}
The units are made out of the masses (which in turn depend on $\Lambda$) to make the above equation dimensionally consistent. Next, we find that the same scaling as in eq \eqref{scaling} is exhibited for the restoration time, see Fig. \ref{tkz}. 
 
\subsection{Saddle dynamics as Liouville quantum mechanics}\label{sec:liou} 
We start with the order parameter equation in the limit, $\Lambda/\sigma \rightarrow \infty$ \ie, eq\eqref{limit}.
Under the change of variables to $\sigma =  \Lambda e^{-y}$
, the first derivative cancels, and the equation becomes
\begin{equation}
\ddot y +  2 \Lambda^2 e^{-2y} \frac{\pi -  y g^2 }{ y g^2} = 0.
\end{equation}
Now if we look at the limit, $y\rightarrow \infty$ (keeping $\sigma = \Lambda e^{-y}$ fixed, such that $g^2 y \gg \pi$), quite surprizingly the above equation becomes independent of $g^2$!   $\ddot{y} - 2\Lambda^2 e^{-2y} =0.$ Redefining, $\phi = -2y $ and analytically continuing, $\tau = i t$, the resulting equation of motion follows from the Euclidean action,  \begin{equation}\label{liouac} S = \int d\tau\, (  \frac{\dot{\phi}^2}{8}  +  \Lambda^2 e^{2 \phi } ). \end{equation} This is the action of Liouville quantum mechanics, which results under the double scaling limit, $\Lambda \rightarrow \infty, \phi \rightarrow -\infty$, with $\sigma = \Lambda e^{ \phi/2}$ fixed. 
 Liouville quantum mechanics captures the dynamics of the zero mode of the Liouville field theory which describes 2D induced gravity in the conformal gauge \cite{Ginsparg:1993is}. Note that the Liouville field $\phi$, being real, we focus only on positive $\sigma$. Using the conventions of \cite{Bagrets_2016}, we look at the wavefunction $\Psi[\phi(\sigma)] = \vev{\phi(\sigma) | k }$, which satisfies the Schr\"odinger equation, derived from the Hamiltonian corresponding to the action \eqref{liouac}:
 \be
\label{schrodinger}
-2 \partial_\phi^2 \Psi_k + \Lambda^2 e^{\phi} \Psi_k = 2k^2 \Psi_k. 
\ee
This has a continuous spectra labelled by $k>0$. The wavefunction which is normalized such that at $\phi \rightarrow -\infty$, the state : $\vev{k|k'} = 2\pi \delta(k-k')$ is given by $\Psi_k = \tfrac{2}{\Gamma(2 i k)} K_{2ik}\left(\sqrt{2}\sigma \right)$.\footnote{ We also demand, $\Psi[\sigma] \rightarrow 0$, for $\sigma \gg \Lambda$. }  Close to $k\rightarrow 0$, $|\Psi_k|^2 = 16 K_0(\sqrt{2}\sigma)^2 k^2 + \mathcal{O}(k^3)$. $K_0(x)$ is known to satisfy the inequality: $ \frac{\sqrt{\pi} e^{-x}}{\sqrt{2(x +a ) }} < K
_0(x) <\frac{\sqrt{\pi} e^{-x}}{\sqrt{2 x  }}$, for $x>0$ and $a\geq 1/4$,\cite{besselk0}. This implies that the $\sigma \rightarrow 0$ configuration dominates at low energies. 
This provides a likely explanation in the \textit{double scaling limit} for the restoration of chiral symmetry. It is interesting to note that the Liouville quantum mechanics also describes the long time Schwarzian dynamics in the SYK model \cite{Bagrets_2016}. 
\vspace{.5cm}

\section{Discussion}\label{sec:conc}
In this work we have implemented smooth global quench of the quartic coupling in the Gross-Neveu model\footnote{and also in the Nambu and Jona-Lasinio model, (see appendix \S\ref{app:njl}) in 2 dimensions.} at zero temperature and in absence of any chemical potential. We find that as a result of the quench, the order parameter, which is the dynamical symmetry breaking fermionic mass, becomes time dependent. Our main finding is that the quench drives the order parameter asymptotically to zero, thus restoring the symmetry. Additionally, we have shown that the time of the symmetry restoration exhibits Kibble-Zurek scaling. Very interestingly, the quench dynamics echoes Liouvillian dynamics which raises further questions. In \S\ref{efftherm} we have also interpreted the dynamical symmetry broken-restoration transition as an \textit{effective thermalization}. In  smooth quenches in the non-linear $O(N)$ model \cite{Gemsheim:2019qed} and in the SYK model\cite{Eberlein:2017wah}, effective thermalization was also observed. The interplay of smooth quenches and thermalization was also studied holographically in \cite{Bhaseen:2012gg}. We have not studied extensively the dependence of the temperature and the thermalization rate on the quench amplitude or the quench rate, which we leave for future work. \\ 
Now, it is known that the thermal GN model (in $2+1$ dimensions) can be mapped to the non-critical M-theory \cite{Petkou:2005se, Horava:2005tt}. The non-critical M-theory contains the $c=1$ Matrix model in its solution space which in the double scaling limit is described by the Liouville conformal field theory \cite{Ginsparg:1993is}. Since we start from a $1+1$ dimensional GN theory, it may not be surprizing that we end up with the Liouville quantum mechanics. It may be insightful to follow this series of connections and understand effective thermalization following a quench, from this perspective. Another situation where similar physics seems to play a role is in \cite{PhysRevB.91.054306},\footnote{We thank A. Polkovnikov for bringing this work to our attention.} wherein one once again solves driven self-consistency equations similar to eq\eqref{saddle2}. Interestingly, the authors find an effective \textit{heating} near the critical point and a KZ scale emerge. Similar ideas of symmetry restoration upon driving also seem to be deep-seated in \cite{Kofman_2004}. Apart from a more extensive understanding of these connections, we also leave a few technicalities for future work.\\
\textbf{Fast ? } In this work we were always in the KZ regime, such that $\delta t^{-1} < m_i, m_f$. Fast scalings emerge when the rate is larger than the mass scales \cite{Das:2014hqa}. Presently our derivative expansion does not allow us to access this regime. A full numerical treatment of the problem by solving directly equation \eqref{ode1} will be necessary to study fast quenches.\\
\textbf{1/N ?} It will also be challenging but useful to investigate the $1/N$ corrections to the saddle point equations. There will be two sources, one coming from the Schwinger-Keldysh propagator's off-diagonal entries, and the other from the traditional $1/N$ fluctuations around the semi-classical saddle. If the quench really thermalizes the system, then the expectation is that the fluctuations will be suppressed. It is to be noted, that using Liouville quantum mechanics \cite{Bagrets_2016} in the double scaling regime, one can estimate the fall-off $\vev{\sigma(t) \sigma(0)} \sim t^{-3/2}$. It will also be interesting to check if the suppression of matrix elements during the quench is consistent with the Eigenstate Thermalization Hypothesis. The ETH gives a likely criteria for thermalization \cite{DAlessio:2016rwt}, note however \cite{Mori:2017dip}. \\
\textbf{Holography ? } A UV completion of the zero temperature Gross-Neveu theory in the string theoretic setting has been formulated in terms of intersecting D4-D6 branes in \cite{Antonyan:2006qy, Basu:2006eb}. In the weak string coupling limit, the 1+1-D Gross-Neveu theory emerges in this set-up where the role of the four-fermion coupling  is played by the inverse of the separation between the D6 branes. Thus a global quench would amount to studying the effective D-brane dynamics under a drive that moves the D6 branes around. It will be interesting to investigate how KZ arises here and what happens in the strong string coupling regime which is when the D-branes are too close to each other. It will also be interesting if the near horizon geometry in this set-up has any semblance of AdS${}_2$ which may give a holographic explanation for the emergence of the Liouville quantum mechanics description. \\ 
\textbf{Chaos ?} Another interesting computation will be that of the out of time ordered correlator \cite{1969JETP...28.1200L, Maldacena:2015waa}, both in the field theory as well as in holography. It will be interesting to compare the KZ time regime with the scrambling time scale. The Lyapunov exponent was also defined recently in the context of evolution of chiral condensates \cite{Hashimoto:2016wme}, it will be interesting to also explore this in the NJL${}_2$ context. \\
\textbf{Quenching in extended phases ?} The Gross-Neveu model also has a rich phase diagram at finite temperature and charge density exhibiting both second as well as a first order transition \cite{Wolff:1985av}. An obvious generalization to the present study is to understand the quenches between phases in this rich set-up. Additionally both the zero charge / temperature as well as the finite case is known to have integrability structures \cite{Basar:2010mu}, which had been used to find spacetime dependent condensate solutions to the gap equation. Understanding how such solutions dynamically contribute during a quench may lead to insights towards understanding the KZ scaling. \\ 
To conclude : There are a lot of interesting non-equilibrium phenomena for which the Gross-Neveu model provides an excellent setting for fruitful investigations. \\

\section*{Acknowledgement}
It is a pleasure to thank Joydeep Chakrabortty, Sumit R. Das, Amit Dutta, Arijit Kundu, R. Loganayagam, Sridip Pal, Anatoli Polkovnikov, and Spenta R. Wadia for discussions and useful comments on the manuscript. The authors will like to acknowledge the support provided by the Max Planck Partner Group grant MAXPLA/PHY/2018577.

\appendix 
\section{Details of the derivative expansion}\label{app:eps}
From equations \eqref{h1h2} and \eqref{ansatz} the equation satisfied by $\omega_k(t)$ is, 
\begin{align}\label{ode1}
&4m^2 (k - \omega_k)^2 \omega_k + 8k \omega_k^4 - 4 i \omega_k^3 \dot\omega_k +\frac{3 k^2 \dot\omega_k^2}{\omega_k } + \omega_k^2 ( 4i k \dot{\omega}_k - 8 k^3 ) \nn \\  &+ \frac{2}{m}   \dot{m} ( k - \omega_k) ( 2 i \omega_k ( k^2 - \omega_k^2) + k \dot{\omega}_k ) 
+ 2 k \omega_k ( \ddot\omega_k + 2k^3 )\nn \\
&= 4 \omega_k^5 + 2k( 2\dot\omega_k^2 + k \ddot{\omega}_k ).
\end{align}
We solve the above equation in an $\epsilon$ expansion, $$ \omega_k = \sum_{n\geq 0} \epsilon^n \omega_k^{(0)}, \,\,\, \partial_t \rightarrow \epsilon \partial_t. $$ To $\mathcal{O}(\epsilon^0 )$ we have,  
\begin{equation}
\omega_k^{(0)} = \sqrt{k^2 + m^2 }. 
\end{equation}
At the next order, the equation yields, $\omega_k^{(1)} = 0$. To the second order in $\epsilon$ we obtain, 
\begin{align}
\omega_k^{(2)} &= \frac{ k \, m^2 \dot{m}^2 ( 6\sqrt{m^2 + k^2 } - 5k )}{ 8 ( 2k (\sqrt{k^2 + m^2 } - k ) - m^2 ) ( k^2 + m^2)^{5/2} } \nn \\
&+   \frac{m\, \ddot{m}\, k \,( k - \sqrt{k^2 + m^2 } ) }{ 4 ( 2k (\sqrt{k^2 + m^2 } - k ) - m^2 ) ( k^2 + m^2)^{3/2} }.
\end{align}
We have checked for self-consistency that $\omega_k^{(2)} > \omega_k^{(0)}$. Note, that what appears in the R.H.S of \eqref{saddle} is the correlator for mass $= -\sigma(t)$ however since, both $\omega_k^{(0)}(t)$ as well as $\omega_k^{(2)}(t)$ are symmetric under $m\rightarrow -m$, we can just replace $m$ with $\sigma$ in $\omega_k(t)$. 
Next we plug $\omega_k(t) = \omega_k^{(0)}(t) + \epsilon^2 \omega_k^{(2)}(t)$ in the integrand of equation \eqref{saddle2} and once again keep till $\mathcal{O}(\epsilon^2)$ terms. This yields,
\begin{eqnarray}\label{Akt}
\mathcal{A}(k,t) &=& \frac{\sigma}{\sqrt{k^2 + \sigma^2 }} + \frac{\dot{\sigma}^2 k^3 \sigma ( 5k - 6 \sqrt{k^2 + \sigma^2} )}{8 \bigg( \sigma^2 + 2k ( k - \sqrt{k^2 + \sigma^2 } )\bigg) \bigg( k^2 + \sigma^2\bigg)^{7/2} } \nn \\
&+& \frac{\ddot{\sigma} k^3 \bigg( k + \sqrt{k^2 + \sigma^2 } \bigg)}{4\sigma^2 \bigg( k^2 + \sigma^2 \bigg)^{5/2} }. 
\eea 

\section{Gap quench, further details}\label{app:tanhnum}
\counterwithin{figure}{section}
We present the details relevant to \S\ref{sec:KZ} for the $\tanh$ quench case of the equilibrium mass gap parameter, $m(t)$. 
We choose our parameters to be, 
\begin{eqnarray}
\label{params}
\Lambda &=& 11, \,\,\, m_i = 0.03, \,\,\, m_f = 0.6, \,\,\, \delta t \,\, \text{to vary from }\,\, 34\, \text{ to }\,\, 65. 
\end{eqnarray}
\begin{figure}[!htbp]
        \centering
                \includegraphics[scale=0.5]{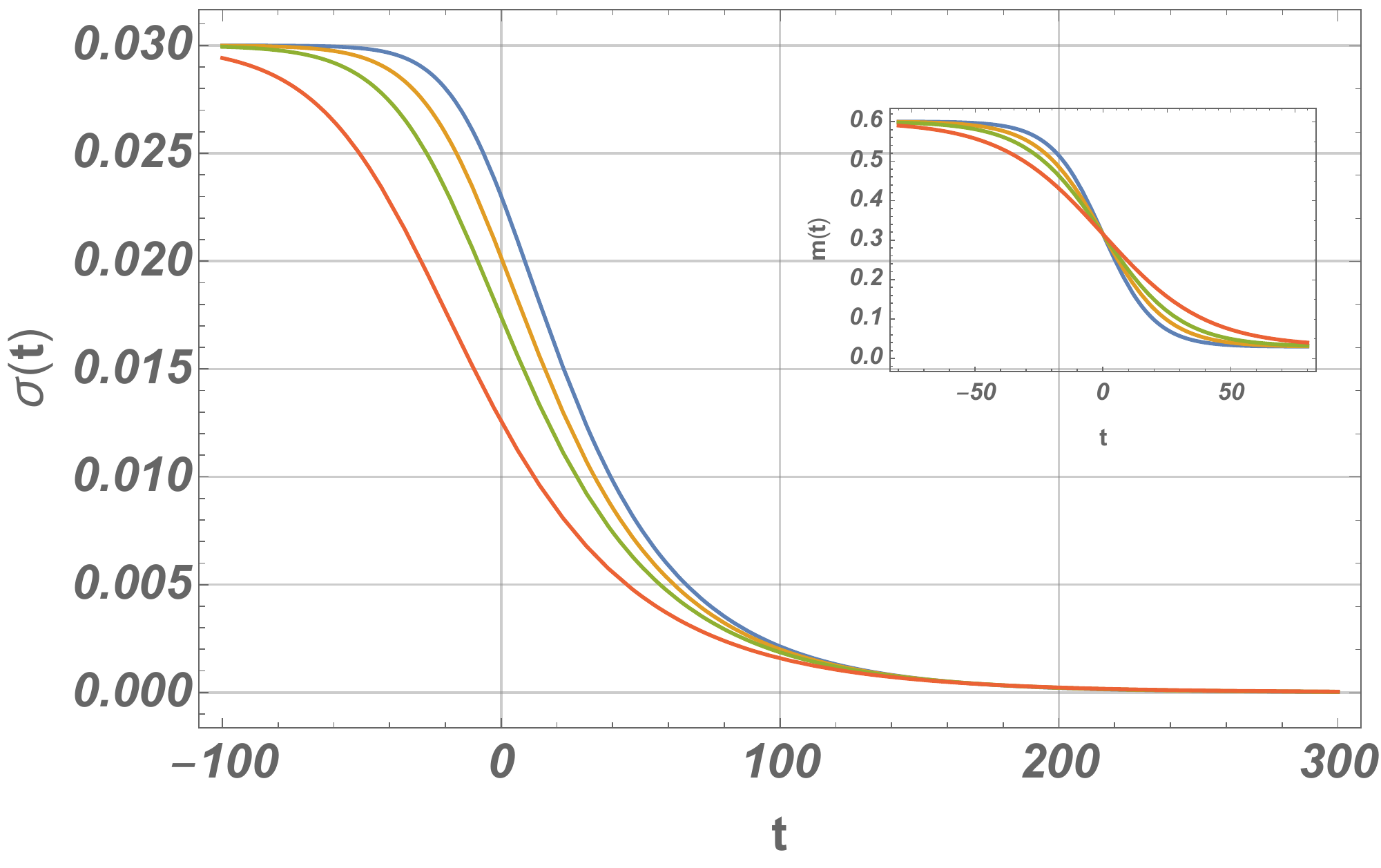} 
                \caption{Plot of $\sigma(t) \text{ vs.}  t$ for the various rates. The inset curve shows the various quench profiles of $m(t)$. The colours correspond to different rates, blue : 1/20, orange : 1/25, green : 1/30 and red : 1/40. }
                \label{tanhfigs}
\end{figure}
\begin{figure}[!htbp]
        \centering
                \includegraphics[scale=0.5]{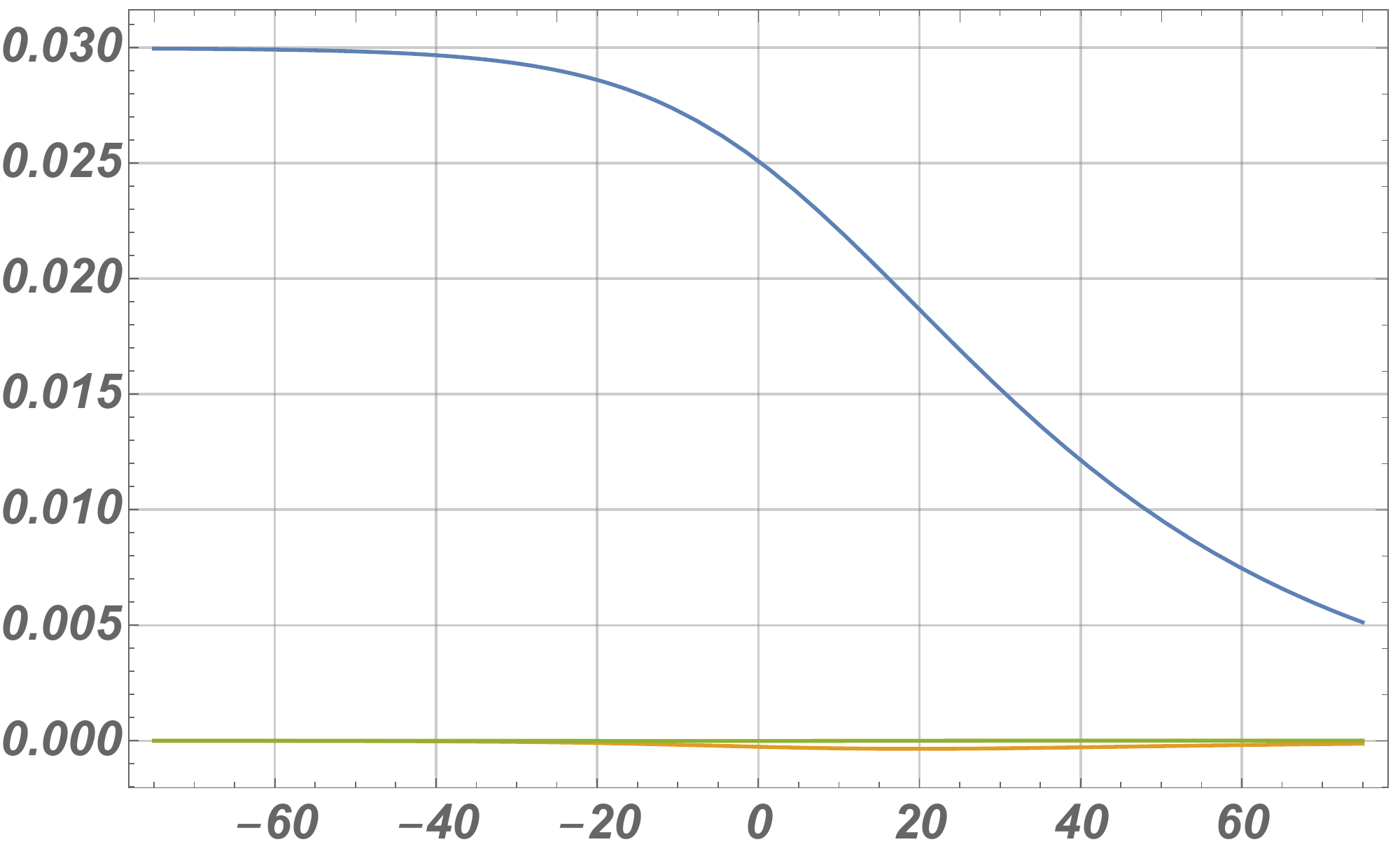} 
                \caption{Plot of $\sigma(t) \text{ and its time derivatives vs.}  t$ for $\Lambda = 11, \,\,\, m_i = 0.03, \,\,\, m_f = 0.6, \,\,\, \delta t = 25$. The blue curve is $\sigma(t)$, while the green and the orange are the first and second derivatives respectively.  }
                \label{b5der}
\end{figure}

As we see in Fig.\ref{tanhfigs} naively it seems that the slower rates have a faster change in the $\sigma(t)$ profiles, however if we zoom in towards the tail we find that there is a crossing, and the $\sigma(t)$ trajectories remain that way henceforth. This is shown in Fig.\ref{cross}.
 \begin{figure}[!htbp]
\centering
\includegraphics[scale=0.5]{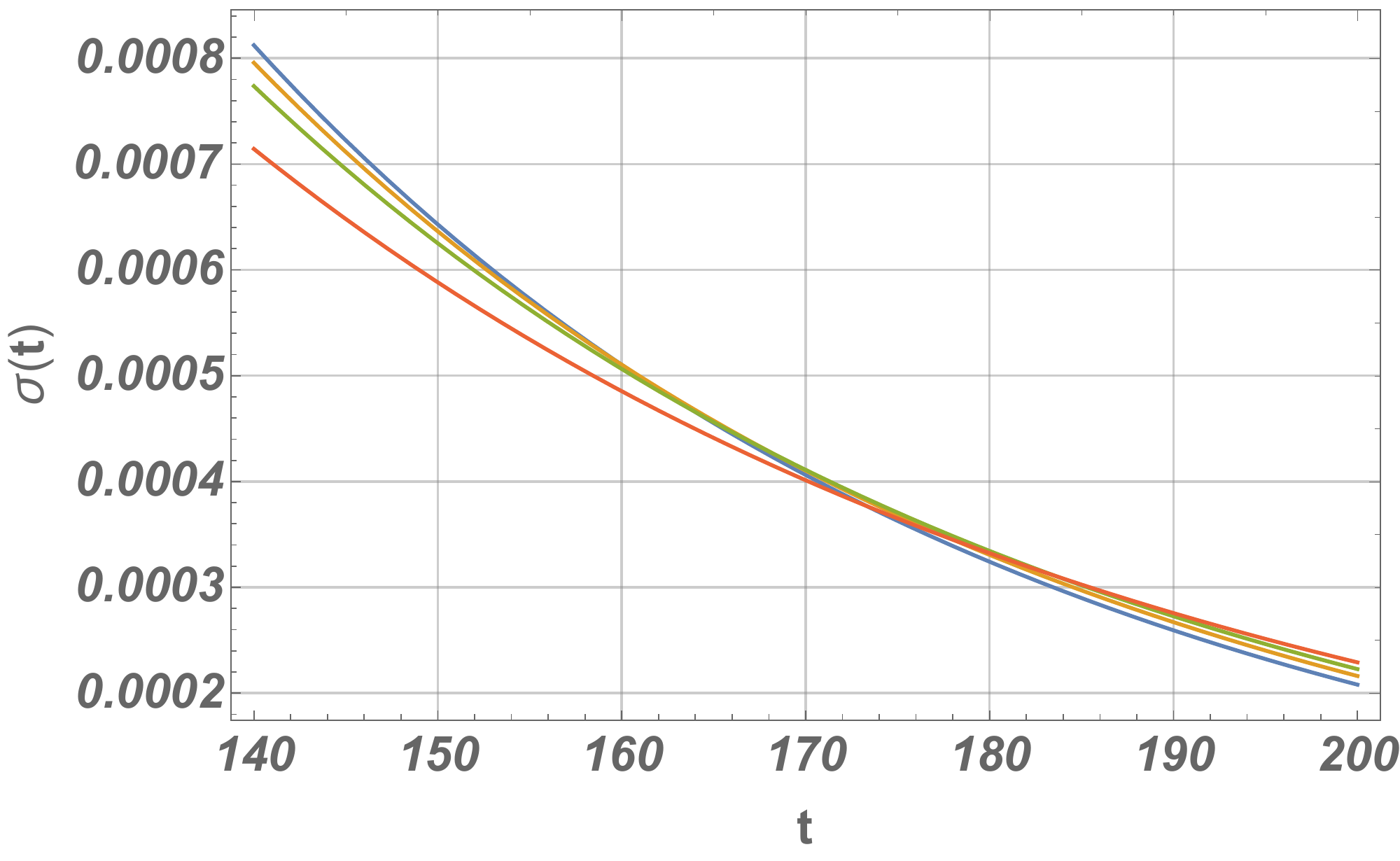} 
\caption{The $\sigma(t)$ trajectories cross and thus the $\sigma(t)$ corresponding to faster rates go to zero earlier.}
\label{cross}
 \end{figure}
 Next we choose a small fixed value, $\epsilon = 0.00001$ and extract the times from the $\sigma(t)$ trajectories using the $\texttt{FindRoot}$ function of $\texttt{Mathematica}^{\tiny{\textregistered}}$. A visual illustration of the process is shown in Fig.\ref{tzero}. 
 \begin{figure}[!htbp]
\centering
\includegraphics[scale=0.5]{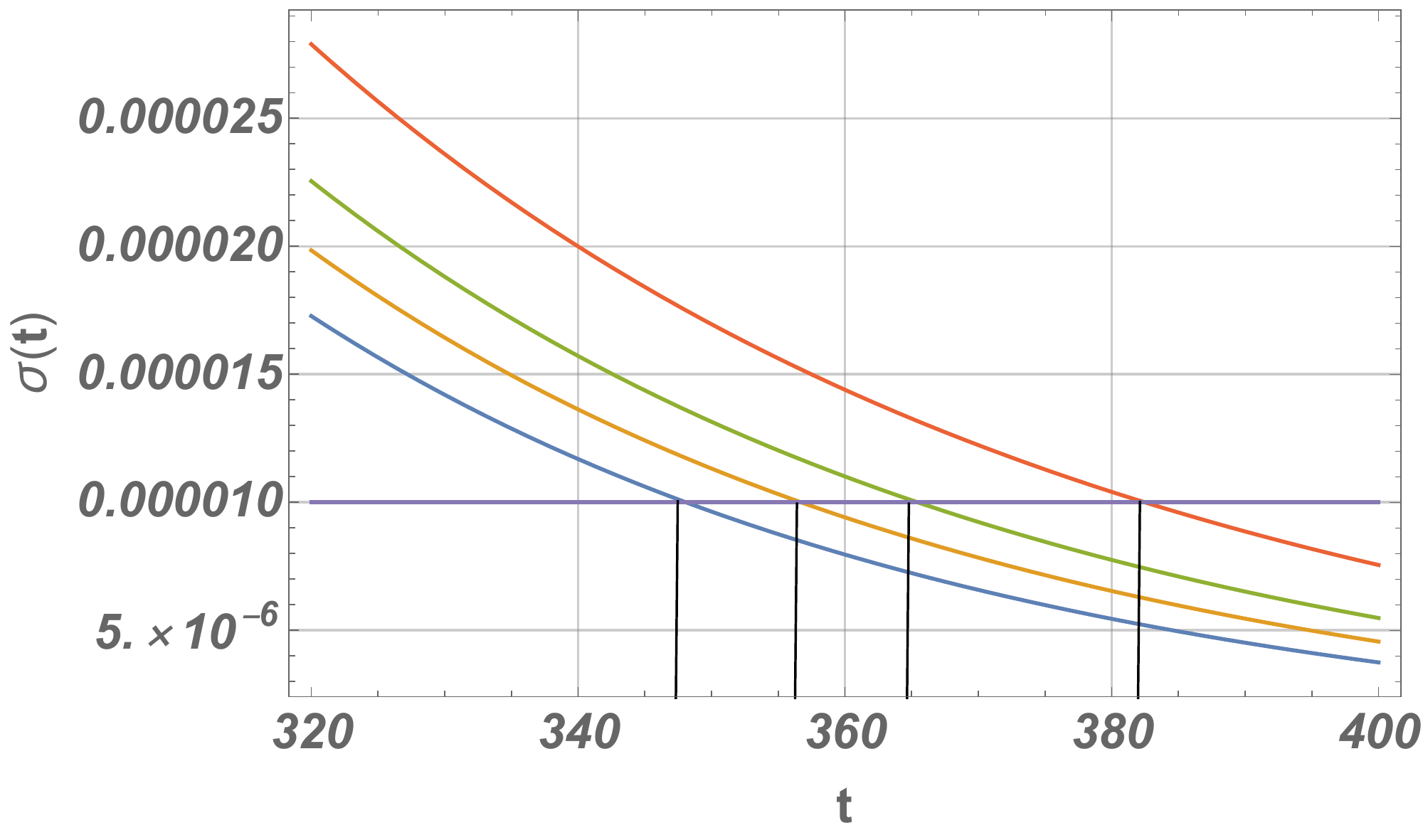}
\caption{The zero times are the places in the $t$-axis where the black straight lines intersect.}
 \label{tzero}
 \end{figure}
 \FloatBarrier
 
\subsection{The Nambu \& Jona-Lasinio model in two dimensions}\label{app:njl}
This is a theory of  N massless interacting fermions \cite{Nambu:1961tp}.
The Lagrangian density of this model is given by,
\begin{equation}
\mathcal{L} = \bar{\psi}_{i} i \slashed{\partial} \psi_{i} + \frac{g^2}{2N} \left[(\bar{\psi}_{i} \psi_{i})^2 -(\bar{\psi}_{i}\gamma_5 \psi_{i})^2 \right], \,\,\, i = 1, 2\dots N.\label{lag1}
\end{equation}
Under chiral transformation  $\psi \rightarrow \exp(i \alpha\gamma_5) \psi$, 
\begin{eqnarray}
\bar{\psi}\psi + \bar{\psi} \gamma_5 \psi &\rightarrow & \exp(2 i \alpha)(\bar{\psi}\psi + \bar{\psi} \gamma_5 \psi), \,\,\,\,
\bar{\psi}\psi - \bar{\psi} \gamma_5 \psi \rightarrow \exp(-2 i \alpha)(\bar{\psi}\psi - \bar{\psi} \gamma_5 \psi)
\end{eqnarray}
So, under a given $U(1)_A $ transformation above Lagrangian remains invariant. This continuous $U(1)_A $ symmetry disallows the theory from generating a mass term. However the symmetry gets broken in the equilibrium and generation of mass occurs. To see this we use similar trick as we did before for the discrete  $\mathbb{Z}_2$ case (Gross-Neveu model). However, here we need to introduce two scalar fields, in terms of which the action can be written as,
\begin{eqnarray}
S=\int d^2 x \{ \bar{\psi} i \slashed{\partial} \psi - \frac{N}{2 g^2} (\sigma^2 + \Pi^2) + (\sigma \bar{\psi} \psi + i \Pi \bar{\psi}\gamma_5 \psi )\} \label{action}
\end{eqnarray}
The equations of motion for $ \sigma$ and $ \Pi$ are respectively,
\begin{eqnarray}\label{eom1}
\sigma &=& \frac{g^2}{N}(\bar{\psi} \psi), \,\,\,\
\Pi = \frac{g^2}{N}(i \bar{\psi}\gamma_5 \psi)
\end{eqnarray}
We can write these two equation of motion in a combined form
\begin{eqnarray}
\sigma \pm i \Pi = \frac{g^2}{N} \left[\bar{\psi}(1 \mp \gamma_5 )\psi \right]
\end{eqnarray}
The action (\ref{action}) remains invariant under $U(1)_A$ provided this combination transforms as :
\begin{eqnarray}
(\sigma + i \Pi) \rightarrow \exp(-2 i\alpha)(\sigma + i \Pi)
\end{eqnarray}
The path integral now becomes, 
\begin{equation}
\mathcal{Z} = \int \mathcal{D} \bar{\psi} \mathcal{D}\psi \mathcal{D}\sigma \mathcal{D}\Pi \exp \left[ i \int d^2 x ( \bar{\psi} i \slashed{\partial} \psi - \frac{N}{2 g^2} (\sigma^2 + \Pi^2) + (\sigma \bar{\psi} \psi + i \Pi \bar{\psi}\gamma_5 \psi ) ) \right] \label{path}
\end{equation}
 If we integrate out the $ \sigma$ and $\Pi$ fields then we get back the original path integral (ignoring some irrelevant constant)
 \begin{eqnarray}
 \mathcal{Z} = \int \mathcal{D} \bar{\psi} \mathcal{D}\psi \exp \left[ i \int d^2 x (\bar{\psi}_{i} i \slashed{\partial} \psi_{i} + \frac{g^2}{2N} \{(\bar{\psi}_{i} \psi_{i})^2 -(\bar{\psi}_{i}\gamma_5 \psi_{i})^2 \})\right]
 \end{eqnarray}
 In equilibrium one can consider $\sigma $ and $\Pi $ as space-time constant fields. Integrating out equilibrium fermions and then evaluating the saddle in the $ \sigma$ and $\Pi$ functional integrals gives, $$
 \sigma^2 + \Pi^2 = \Lambda^2 \exp(-\frac{2\pi}{g^2}).$$ 
 Now if we parameterise the complex scalar fields as $\sigma + i \Pi = \rho e^{i \theta} $, then we have $
 \rho = \Lambda \exp(-\frac{\pi}{g^2})$, 
 where $ \Lambda $ is cut-off scalar. For non-zero coupling $ \rho$ has some non-zero value which is a signal of $U(1)_A$ symmetry breaking and as a result we get a mass term for the fermions.
 Now for implementation of quench in this model we once again promote $ g \rightarrow g(t)$. Therefore all amplitudes need to be calculated using the Schwinger-Keldysh prescription and $\sigma$ and $\Pi$ become time dependent fields.\\
The partition function in Schwinger-Keldysh formalism can be expressed as,
\begin{eqnarray}
\mathcal{Z} = \int \mathcal{D} \bar{\psi}_{\pm} \mathcal{D}\psi_{\pm} \mathcal{D}\sigma_{\pm} \mathcal{D} \Pi_\pm \exp \left[ i \{S(\bar{\psi}_{+},\psi_{+}, \sigma_{+}) - S(\bar{\psi}_{-},\psi_{-}, \sigma_{-}) \} \right] 
\end{eqnarray}

where, we have
\begin{align}
S(\bar{\psi}_{+},\psi_{+}, \sigma_{+}) - S(\bar{\psi}_{-},\psi_{-}, \sigma_{-}) = \int d^2 x [\bar{\psi}_{+} ( i  \slashed{\partial}  + \sigma_{+}+ i \Pi_+ \gamma_5 ) \psi_{+}\nonumber \\ - \bar{\psi}_{-} ( i \slashed{\partial} + \sigma_{-} + i \Pi_- \gamma_5 ) \psi_{-} \nonumber  - \frac{N}{2 g^2(t)} (\sigma^2_{+} - \sigma^2_{-}) - \frac{N}{2 g^2(t)} (\Pi^2_{+} - \Pi^2_{-}) ].
\end{align}
 After integrating out $\psi_\pm $ and $\bar{\psi}_\pm$ we have, the effective action, 
 \begin{eqnarray}
 S_{eff} = -N\,\text{Tr} \log D - \frac{N}{2 g^2(t)} \int d^2 x (\sigma^2_{+} - \sigma^2_{-}) \nonumber \\- \frac{N}{2 g^2(t)} \int d^2 x(\Pi^2_{+} - \Pi^2_{-})\label{seff}
 \end{eqnarray}
where, 
\begin{equation}
D= \left( \begin{matrix}
i\slashed{\partial} + \sigma_{+} + i \gamma_5 \Pi_+   && 0 \\
0 								&& - i\slashed{\partial} - \sigma_{-} - i \gamma_5 \Pi_-  
\end{matrix} \right).
\end{equation}
The large-$N$ saddle equation is, 
\begin{eqnarray}\label{sadds}
\frac{\partial S_{eff}}{\partial \sigma_\pm} =0 \nonumber \\
\Rightarrow \frac{\sigma_\pm}{g^2}= - \text{Tr} \left[ \frac{1}{i  \slashed{\partial}  + \sigma_{\pm} + i \Pi_\pm \gamma_5} \right]\label{eom2}\\
\frac{\partial S_{eff}}{\partial \Pi_\pm} =0 \nonumber \\
\Rightarrow \frac{\Pi_\pm}{g^2}= - i \text{Tr} \left[ \frac{\gamma_5}{i  \slashed{\partial}  +  \sigma_{\pm} + i \gamma_5 \Pi_\pm } \right]\label{eom3}
\end{eqnarray} 
The R.H.S's are fermionic propagators of a theory with the quadratic term in the action : $\bar{\chi}(i  \slashed{\partial}  +  \sigma_{\pm} + i \gamma_5 \Pi_\pm)\chi $.\\
In our convention: 
$
\gamma^5 = \begin{pmatrix}
-1 && 0 \\ 0 && 1
\end{pmatrix}.$
In the momentum space the equations \eqref{sadds} become
\begin{align}
&\frac{\sigma(t)}{g^2(t)} = -\int dk \langle \bar{\chi}_{k,1}\chi_{k,1}\rangle , \,\,\,\,
\frac{\Pi(t)}{g^2(t)} = - i \int dk \langle \bar{\chi}_{k,2}\gamma_5\chi_{k,2}\rangle \label{eompisigma}
\end{align}
where these fermions satisfy ,
\begin{eqnarray}
\begin{pmatrix}
\sigma - i \Pi && i(\partial_{t} + \partial_{x}) \\
i(\partial_{t} - \partial_{x}) && i \Pi + \sigma
\end{pmatrix}\chi_j(x,t) = 0\label{eomchi2} \,\,\text{with }\, j =1,2.
\end{eqnarray}
Now we can proceed with the spinor ansatz:
\begin{align}
u_{k,j}(x,t) &= \frac{ e^{ikx} }{\sqrt{2\pi }  } \begin{pmatrix} h_{k,j}^I(t) \\ -h_{k,j}^{II}(t) \end{pmatrix} , \,\,\,
v_{k,j}(x,t) = \frac{ e^{-ikx} }{\sqrt{2\pi }  } \begin{pmatrix} h_{-k,j}^{II*}(t) \\ h_{-k,j}^{I*}(t) \end{pmatrix}, \,\,\, \text{ for $j = 1,2$ }.  \label{uv}
\end{align}
The equations of motion in momentum space is now the following coupled ODEs,
\begin{eqnarray}
\dot{h}_{k,j}^{II}(t) + i k h_{k,j}^{II}(t) + (\Pi + i \sigma)h_{k,j}^I(t) =0 \label{hII} \\
\dot{h}_{k,j}^{I}(t) - i k h_{k,j}^{I}(t) -(\Pi - i \sigma)h_{k,j}^{II}(t) =0 \label{hI}
\end{eqnarray}
along with the standard normalization condition,$| h_{\pm k,j}^{I} |^2 + | h_{\pm k,j}^{II} |^2 =1 $.
Next, we write the second-quantized mode expansions of the Dirac fields as:
\begin{eqnarray}\label{mode}
\chi_{1}(x,t) =\int dk [B_{k,1} u_{k,1}(x,t) + D_{k,1}^{\dagger} v_{k,1}(x,t) ]\\
\chi_{2}(x,t) =\int dk [B_{k,2} u_{k,2}(x,t) + D_{k,2}^{\dagger} v_{k,2}(x,t) ]
\end{eqnarray}
with, $ \{B_{k,j},B^{\dagger}_{k',j} \} = \delta(k-k') $ and $\{D_{k,j},D^{\dagger}_{k',j} \} = \delta(k-k') $. One can check easily that, $ \{\chi_{j}(x,t),\chi^{\dagger}_{j}(x',t) \}= \delta(x-x')$.
Using the commutators we obtain from equations (\ref{eompisigma}),
\begin{eqnarray}
\frac{\sigma(t)}{g^2(t)} = -\int \frac{d k}{2 \pi} [h_{-k,1}^{I} h_{-k,1}^{II*} + h_{-k,1}^{II} h_{-k,1}^{I*} ]\label{f1}
\end{eqnarray}
 and,
 \begin{eqnarray}
\frac{\Pi(t)}{g^2(t)} = i \int \frac{d k}{2 \pi} [h_{-k,2}^{I} h_{-k,2}^{II*} - h_{-k,2}^{II} h_{-k,2}^{I*} ]. \label{f2}
\end{eqnarray}

To proceed we propose the following \textit{ansatz}, 
\begin{eqnarray}
h_{k,j}^I &=& \sqrt{ \frac{\omega_k(t) - k }{2\omega_k(t) }} e^{-i \int^t \omega_k(t') dt' } \nonumber \\
 h_{k,j}^{II} &=& - \frac{\sqrt{\sigma^2(t) + \Pi^2(t)}}{\sigma(t) + i \Pi}\sqrt{ \frac{\omega_k(t) + k }{2\omega_k(t) } }e^{-i \int^t \omega_k(t') dt' } \,\,
 \text{for  } j=1,2,
 \label{ansatz2}
\end{eqnarray}
where in equilibrium  $\omega_{k}= \sqrt{k^2 + \Pi^2 +\sigma^2}$. 
Now in terms of  $ \sigma(t) + i \Pi(t) = \rho(t) e^{i \theta(t)} $ and using the \textit{ansatz} one can gets by some algebra

 \begin{align}
 \frac{\rho(t) \cos \theta(t)}{g^2(t)} =&  \int \frac{d k}{2 \pi} \Big[\sqrt{\frac{\omega_k(t) - k}{2 \omega_k(t)}} \left(\sqrt{\frac{\omega_k(t) + k}{2 \omega_k(t)}}\right)^* e^{i \theta(t)} \nonumber \\+& \sqrt{\frac{\omega_k(t) + k}{2 \omega_k(t)}} \left( \sqrt{\frac{\omega_k(t) - k}{2 \omega_k(t)}}\right)^* e^{-i \theta(t)} \Big] \label{1st}\\
 \frac{\rho(t) \sin \theta(t)}{g^2(t)} =&  i \int \frac{d k}{2 \pi} \Big[\sqrt{\frac{\omega_k(t) - k}{2 \omega_k(t)}} \left(\sqrt{\frac{\omega_k(t) + k}{2 \omega_k(t)}}\right)^* e^{i \theta(t)} \nonumber\\-& \sqrt{\frac{\omega_k(t) + k}{2 \omega_k(t)}} \left( \sqrt{\frac{\omega_k(t) - k}{2 \omega_k(t)}}\right)^* e^{-i \theta(t)} \Big]. \label{2nd}
  \end{align}
 If $\omega_{k}$ is real then these two equations reduce to a single equation
 \begin{equation}
 \frac{\rho(t)}{g^2(t)} = \int \frac{d k}{2 \pi} \frac{\sqrt{\omega_k^2 - k^2}}{\omega_k}
 \end{equation}

 But in general equations (\ref{1st}) and (\ref{2nd}) can be written as
\begin{align}
\frac{\rho(t) \cos \theta(t)}{g^2(t)} =& \int \frac{d k}{2 \pi} \Big[\sqrt{\frac{|\omega_k(t)|^2 - k^2 + 2i \, k \, \text{Im} \omega_{k}(t)}{4 |\omega_k(t)|^2}}  e^{i \theta(t)}  \nonumber\\ +& \sqrt{\frac{|\omega_k(t)|^2 - k^2 - 2i \, k \, \text{Im} \omega_{k}(t)}{4 |\omega_k(t)|^2}} e^{-i \theta(t)} \Big] 
\end{align} 

 \begin{align}
 \frac{\rho(t) \sin \theta(t)}{g^2(t)} =& i\int \frac{d k}{2 \pi} \Big[\sqrt{\frac{|\omega_k(t)|^2 - k^2 + 2i \, k \, \text{Im} \omega_{k}(t)}{4 |\omega_k(t)|^2}}  e^{i \theta(t)} \nonumber \\-& \sqrt{\frac{|\omega_k(t)|^2 - k^2 - 2i \, k \, \text{Im} \omega_{k}(t)}{4 |\omega_k(t)|^2}} e^{-i \theta(t)} \Big] 
\end{align}  
 These last two equations are actually equivalent to the equation :
 \begin{align}
 \frac{\rho(t)}{g^2(t)} =&2 \int_{0}^{\Lambda} \frac{d k}{\pi} \left[\sqrt{\frac{|\omega_k(t)|^2 - k^2 + 2i \, k \, \text{Im} \omega_{k}(t)}{4 |\omega_k(t)|^2}} \right] \label{final} \\
 =& 2\int_{0}^{\Lambda} \frac{d k}{\pi} \mathcal{A}(k,t) \nonumber
\end{align}
We will solve for $\omega_k(t)$ and then plug into the above. From equations (\ref{hII}) and (\ref{hI}) the equation satisfied by $\omega_{k}$ is 
\begin{align*}
-&2 \omega  (k-\omega ) (2 k^2 \omega -i k \omega'-2 \omega ^3) (\Pi '-i \sigma')+i \Pi(2 k   \omega^2 (2 k (k^2+\sigma^2)+\omega'')\\+& \omega^3 (-8 k (k^2+\sigma^2)+4 i k \omega')+3 k^2 \omega '^2-2 k \omega (k \omega ''+2 \omega '^2)+8 k \omega^5+4 \omega^4 (\sigma^2-i   \omega ')-4 \omega^6) \\+&\sigma (2 k \omega ^2 (2 k^3+\omega '')+\omega^3 (-8 k^3+4 i k   \omega ')+3 k^2 \omega '^2-2 k \omega (k \omega ''+2 \omega '^2)+8 k \omega ^5-4 i \omega ^4 \omega   '-4 \omega^6)\\ +& 4 \Pi ^2 \sigma \omega ^2 (k-\omega )^2+4 i \Pi ^3 \omega^2 (k-\omega )^2+4 \sigma^3   \omega ^2 (k-\omega)^2 = 0.
   \end{align*}
 We solve this equation by $ \epsilon $-expansion ,$$\omega_{k} = \sum_{n \geq 0} \epsilon^{(n)} \omega_{k}^{(n)} .$$
 In terms of $\sigma + i\, \Pi = \rho e^{i\, \theta}$ we get solution of $ \omega_{k}$ (up-to 2nd order):
\begin{align*}
\omega_{k}(t) =& \frac{1}{2} \left(-\frac{k}{\sqrt{k^2+\rho^2}}-1\right) \theta'+\sqrt{k^2+\rho^2} \\-& \rho \Big[\frac{k \rho \left(2 k^2
   \left(k-\sqrt{k^2+\rho^2}\right) \theta'^2+\left(5 k-6 \sqrt{k^2+\rho^2}\right) \rho'^2\right)}{8 \left(k^2+\rho^2\right)^{5/2} \left(2 k \left(\sqrt{k^2+\rho^2}-k\right)-\rho^2\right)}\Big] \\-& \rho\Big[\frac{k \rho^3 \left(3 k-2   \sqrt{k^2+\rho^2}\right) \theta'^2+2 k \rho^2 \left(\sqrt{k^2+\rho^2}-k\right) \rho ''+2 k^3 \left(\sqrt{k^2+\rho
   ^2}-k\right) \rho''+\rho^5 \theta '^2}{8 \left(k^2+\rho^2\right)^{5/2} \left(2 k \left(\sqrt{k^2+\rho
   ^2}-k\right)-\rho^2\right)}\Big] \\+& \,i\, \Big[  -\frac{\rho \left(\rho \left(k^2+\rho^2\right) \theta''-2 k \sqrt{k^2+\rho^2} \theta' \rho'+\rho^2 \left(-\theta'\right) \rho'\right)}{4 \left(k^2+\rho^2\right)^{3/2} \left(2 k \left(\sqrt{k^2+\rho^2}-k\right)-\rho^2\right)} \Big].
\end{align*} 
 
 Next putting this $ \omega_{k}(t)$ in the integrand of (\ref{final}) and keeping up to $ \mathcal{O}(\epsilon^2)$ term we get,
  \begin{scriptsize}
 \begin{align*}
\mathcal{A}(k,t) =& \frac{8 k \rho^7 \left(2 \sqrt{k^2+\rho^2}-5 k\right)-16 k^7 \rho \left(k-\sqrt{k^2+\rho^2}\right)+8 k^5 \rho^3 \left(6
   \sqrt{k^2+\rho^2}-7 k\right)+24 k^3 \rho^5 \left(2 \sqrt{k^2+\rho^2}-3 k\right)-8 \rho^9}{16 \left(k^2+\rho^2\right)^{7/2}
   \left(2 k \left(\sqrt{k^2+\rho^2}-k\right)-\rho^2\right)} \\ 
   + & \frac{4 k^2 \rho^5 \left(\sqrt{k^2+\rho^2}-k\right) \theta'-4 k^6 \rho \left(k-\sqrt{k^2+\rho^2}\right) \theta'+ 8 k^4
   \rho^3 \left(\sqrt{k^2+\rho^2}-k\right) \theta'}{16 \left(k^2+\rho^2\right)^{7/2} \left(2 k \left(\sqrt{k^2+\rho^2}-k\right)-\rho^2\right)} \\ +& \frac{2 k^2 \rho^5 \theta'^2 - k^5 \rho \left(k-2 \sqrt{k^2+\rho^2}\right) \theta'^2 + k^3 \rho^3 \left(2 \sqrt{k^2+\rho^2}+k\right) \theta'^2 - k^3 \rho\left(5 k - 6 \sqrt{k^2+\rho^2}\right) \rho'^2}{16 \left(k^2+\rho^2\right)^{7/2} \left(2 k \left(\sqrt{k^2+\rho^2}-k\right)-\rho^2\right)} \\ +& \frac{2 k^5 \left(k-\sqrt{k^2+\rho^2}\right) \rho'' + 2 k^3 \rho^2 \left(k-\sqrt{k^2+\rho^2}\right) \rho''}{16 \left(k^2 + \rho^2\right)^{7/2} \left(2 k \left(\sqrt{k^2+\rho^2}-k\right)-\rho^2\right)} \\ +&  i \Big[\frac{-2  k \rho^5 \sqrt{k^2+\rho^2} \theta''+2 k \rho^4 \left(\sqrt{k^2+\rho^2}+2 k\right) \theta' \rho'-4  k^3 \rho^3 \sqrt{k^2 + \rho^2} \theta''}{16 \left(k^2 + \rho^2\right)^{7/2} \left(2 k \left(\sqrt{k^2+\rho^2}-k\right)-\rho^2\right)} \Big] \\+& i\,\Big[ \frac{4  k^6 \theta' \rho' - 2  k^5 \rho \sqrt{k^2+\rho^2} \theta'' + 2 k^3 \rho^2 \left(\sqrt{k^2+\rho ^2} + 4 k\right)\theta' \rho'}{16 \left(k^2+\rho^2\right)^{7/2} \left(2 k \left(\sqrt{k^2+\rho^2}-k\right)-\rho^2\right)}\Big].
 \end{align*}
 \end{scriptsize}
Integrating this over momentum with a cut-off $ \Lambda $ we get ,

\begin{align}
- \frac{8 \pi \rho}{g^2} =& \frac{4 i \Lambda \theta' \left(\Lambda +\sqrt{\Lambda^2+\rho^2}\right) \rho'}{\rho^4}+\theta' \left(-4 \tan^{-1}\left(\frac{\Lambda }{\rho}\right)+\frac{i \rho'}{\Lambda^2+\rho^2}\right)-\frac{\rho'^2}{\Lambda ^2 \rho + \rho^3} \nonumber \\ +&\rho \Big(\frac{4 \theta' \left(\Lambda ^2 + \rho^2\right)^2 + \theta'^2 \left(\Lambda +2 \sqrt{\Lambda ^2+\rho^2}\right) \left(\Lambda ^2+\rho^2\right)+\left(\Lambda +\sqrt{\Lambda ^2+\rho^2}\right) \rho'^2}{\left(\Lambda ^2+\rho^2\right)^{5/2}} \nonumber \\-& 8 \log \left(\Lambda +\sqrt{\Lambda^2+\rho^2}\right)\Big)+\frac{-\frac{4 \Lambda  \theta'^2}{\sqrt{\Lambda ^2+\rho^2}} + 4 \theta' \sqrt{\Lambda ^2+\rho^2}+4 \Lambda  \theta' }{\rho} \nonumber \\ +& \frac{ \left(\theta'^2 + i\, \theta''\right) \left(\log \left(\Lambda ^2+\rho^2\right)+2 \log \left(\Lambda +\sqrt{\Lambda ^2+\rho^2}\right)\right)-\frac{4 \Lambda  \rho'^2}{3 \left(\Lambda ^2+\rho^2\right)^{3/2}}}{\rho} \nonumber \\+&\frac{-2 i \Lambda  \theta'' \left(\Lambda +\sqrt{\Lambda ^2 + \rho^2}\right)+\Lambda  \theta'^2 \left(\Lambda +\sqrt{\Lambda ^2+\rho^2}\right)}{\rho^3} \nonumber\\+& \frac{\rho'^2 \left(-\frac{5 \Lambda }{3 \sqrt{\Lambda ^2+\rho^2}}+\log \left(\Lambda ^2 + \rho^2\right)+2 \log \left(\Lambda +\sqrt{\Lambda^2+\rho^2}\right)\right)}{\rho^3} \nonumber \\ -& \frac{8 \Lambda  \rho'' - 3 i \sqrt{\Lambda ^2+\rho^2} \left(\log \left(\Lambda ^2 + \rho^2\right) + 2 \log \left(\Lambda +\sqrt{\Lambda^2 + \rho ^2}\right)\right) \left(2 \theta' \rho' - i \rho''\right)}{3 \rho^2 \sqrt{\Lambda ^2 + \rho^2}} \nonumber \\-& \frac{\rho''}{\Lambda ^2+\rho^2} +\frac{2 \Lambda  \rho''}{3 \left(\Lambda ^2+\rho^2\right)^{3/2}}+ \frac{\rho \rho'' (4 \log (\rho ) + 1)- 8 \rho ^3 \theta' -4 \rho '^2 \log (\rho )+8 \rho ^4 \log (\rho)}{\rho^3} \nonumber \\+ & \frac{ i \rho \theta' \rho ' (8 \log (\rho ) - 1)) -2 \rho^2  \left(\theta '^2 (2 \log (\rho)+1)+2 i \theta '' \log (\rho )\right)}{\rho^3}.\label{difrho}
\end{align}

 Note that the right hand side has both real and imaginary parts. We solve this equation numerically with initial conditions that depend on the quench protocol. When $ \rho $ and $ \theta $ are independent of time we get the equilibrium solution,
$ \rho = \frac{2 e^{\frac{\pi }{g^2}} \Lambda }{e^{\frac{2 \pi }{g^2}}-1}.$
 We take this as initial boundary condition along with $\dot{\rho}(-\infty)=0$, $\theta(-\infty)=0$ and $\dot{\theta}(-\infty)=0$ while we solve equation (\ref{difrho})
\begin{figure}[h!]
        \centering
        \subfigure[Restoration of $U(1)$ symmetry]{
                \includegraphics[scale=0.4]{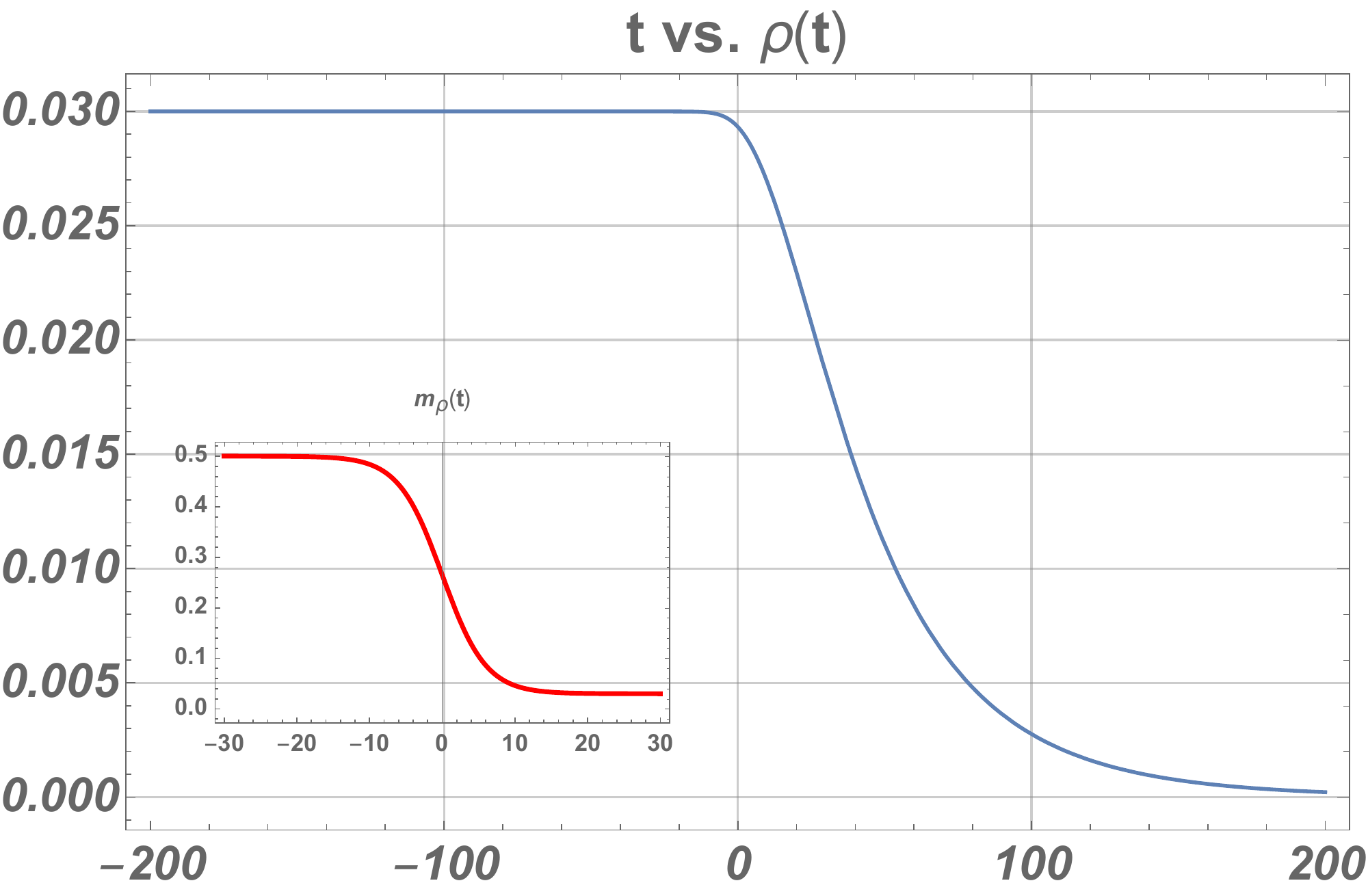} }
         \subfigure[Breaking of $CP$ invariance]{
                \includegraphics[scale=0.4]{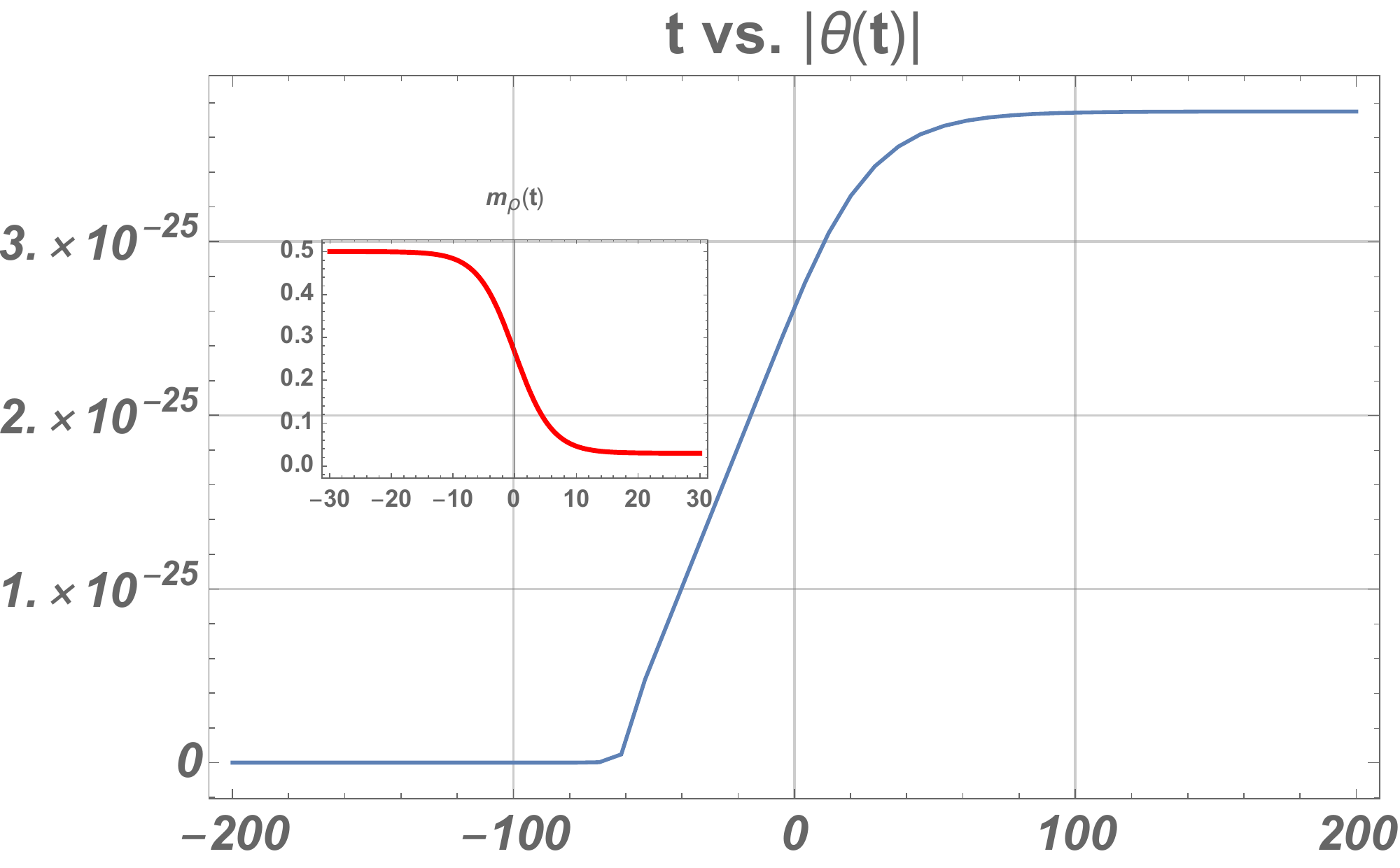} }
                 \caption{(a) Plot of $\rho(t)$ as a function of time. We have. In (b) we plot $\theta(t)$ as a function of time. }
\label{tanh11}
\end{figure}
 The plot in Fig. \ref{tanh11} that order parameter $\rho$ goes to zero after some time which is a signal of the restoration of dynamically broken $U(1)$. We also note that $\theta$ acquired some VEV after some time which is manifestation of CP violation\cite{Grosser:1983jh}. This is consistent with the quench which breaks T and the theory respects CPT \footnote{We thank J. Chakrabortty for pointing this out to us.}.

\begin{footnotesize}

\providecommand{\href}[2]{#2}

\end{footnotesize}
\vspace{2cm}

\end{document}